\documentclass[12pt,reqno]{amsart}
\usepackage{amssymb}
\usepackage{amsthm}
\usepackage{xspace}
\usepackage{amsmath}
\usepackage{setspace}
\usepackage{amscd}
\usepackage{natbib}
\usepackage{setspace}
\usepackage{enumerate}
\usepackage[english]{babel}
\usepackage{tikz}
\usepackage{amssymb}
\usepackage{graphicx}
\usepackage{enumerate}
\usepackage{algorithmic,algorithm}
\usepackage{latexsym, amsmath, amscd, amssymb}
\usetikzlibrary{matrix}
\tikzstyle{l} = [draw, -latex,thick]
\newtheorem{lemma}{Lemma}[section]
\newtheorem{theorem}[lemma]{Theorem}
\newtheorem{corollary}[lemma]{Corollary}
\newtheorem{proposition}[lemma]{Proposition}
\theoremstyle{definition}
\newtheorem{definition}[lemma]{Definition}
\newtheorem{conjecture}[lemma]{Conjecture}

\newtheorem{example}[lemma]{Example}

\theoremstyle{remark}

\begin{document}
\title[ Ideal Lattices and Generalized Hash Functions]
{On Ideal Lattices, Gr\"obner Bases and Generalized Hash Functions}
\author[Maria Francis and  Ambedkar Dukkipati]{Maria Francis and  Ambedkar Dukkipati}
\email{mariaf@csa.iisc.ernet.in \\ ad@csa.iisc.ernet.in}
\address{Department of Computer Science \& Automation\\Indian Institute of Science, Bangalore - 560012}
\maketitle
\begin{abstract}  
In this paper, we draw connections between ideal lattices and
 multivariate polynomial rings over integers using Gr\"obner bases.  
Univariate ideal lattices are ideals in the residue class ring, $\mathbb{Z}[x]/\langle f
\rangle$ (here $f$ is a monic polynomial) and
cryptographic primitives have been built based on these
objects.
Ideal lattices in the univariate case are generalizations of cyclic lattices. We introduce the notion of
multivariate cyclic lattices and show that  ideal
lattices are a generalization of them in the multivariate case too.
Based on multivariate ideal lattices, we construct hash functions
using Gr\"obner
basis  techniques.  We  define a worst case problem,
shortest substitution problem w.r.t. an ideal in $\mathbb{Z}[x_1,
\ldots, x_n]$, and use its computational hardness to establish the collision resistance of the hash functions.
\end{abstract}

\section{Introduction}
\label{Introduction} 
\noindent Ideals in the residue class ring, $\mathbb{Z}[x]/\langle f \rangle$ for any monic polynomial $f \in \mathbb{Z}[x]$, 
are integer lattices as well and hence  are known as ideal lattices.  
This is because $\mathbb{Z}[x]/\langle f \rangle$ is
isomorphic to ${\mathbb{Z}}^{N}$ (as a $\mathbb{Z}$-module) if and
only if $f$ is 
monic. The presence of both ideal and lattice properties make ideal lattices a powerful tool in lattice based cryptography.  The reason why ideal lattices are popular in lattice cryptography is because they provide a compact representation for integer lattices.
In fact, ideal lattices have been used to build several  
cryptographic primitives that include digital signatures \citep{Micciancio:2008:EfficientSign},
hash functions \citep{Lyubashevsky:2006:Ideallatticefirstdef} and  identification schemes \citep{Lyubashevsky:2008:Ideallattice}. Unfortunately, ideal lattices have not been 
studied much outside the periphery of lattice cryptography.
  
After \cite{Ajtai:1996:Znascryptoprimitive} built functions
that on an average generated hard instances of standard lattice
problems, research progressed in the direction of building
cryptographic primitives based on them. 
The fundamental challenge to this direction of research was describing
lattices as $n \times n$ integer matrices,  since that meant the size of
the key and the computation time of the cryptographic functions will 
be atleast quadratic in $n$. 
\cite{Micciancio:2002:CyclicLattices} introduced a class of lattices called `cyclic lattices' to remedy this problem and built certain
efficient one-way functions called generalized compact knapsack
functions using them.
But one way functions are of
theoretical interest and  \cite{Lyubashevsky:2006:Ideallatticefirstdef} introduced the class of ideal lattices, 
which not only gave a succinct representation for lattices but was also a practical tool in building cryptographic primitives. In this paper, we look at how to extend ideal lattices to the multivariate polynomial ring, $\mathbb{Z}[x_1, \ldots, x_n]$.

In algebra, extensions of solutions of problems from the one
variable case to 
the multivariate case have led to important
theories, an  example being the theory of Gr\"obner bases \citep{Buchberger:1965:thesis} 
which is a generalization of the Euclidean polynomial division algorithm in $\Bbbk[x]$. Gr\"obner bases
have since then become a standard tool in computational
algebra and algebraic geometry.  
We show that in the study of multivariate ideal lattices  the theory of Gr\"obner bases plays an important role.
We give a condition for residue
class polynomial rings over $\mathbb{Z}$ to have ideal lattices in
terms of `short reduced Gr\"{o}bner bases' ~\citep{FrancisDukkipati:2013:freeZmodule}. 
We also establish the existence of collision
resistant generalized hash functions based on multivariate ideal
lattices.

\subsection*{Contributions}
Given an ideal $\mathfrak{a}$ in $\mathbb{Z}[x_1,\ldots,x_n]$, we
study the cases for which ideals in
$\mathbb{Z}[x_1,\ldots,x_n]/\mathfrak{a}$ are also lattices. First, we define 
cyclic lattices in the multivariate case.  
We then show that multivariate ideal lattices are a generalization of
multivariate cyclic lattices.  
We show that ideal lattices exist only when the residue class
polynomial ring over $\mathbb{Z}$ is a free $\mathbb{Z}$-module, for
which we give a characterization based on short reduced Gr\"obner
bases  ~\citep{FrancisDukkipati:2013:freeZmodule}.
For the construction of many cryptographic primitives, full rank lattices are essential and we derive the condition for a multivariate  ideal lattice to be full rank. We also give an example
of a class of binomial ideals in $\mathbb{Z}[x_1,\ldots,x_n]$, that gives rise to full rank integer lattices.
To show the existence of collision resistant hash functions, we define an expansion
factor  w.r.t. each variable to accommodate the growth of
coefficients.  We extend
the smallest polynomial problem ($SPP$) for multivariate ideal
lattices. An important result of this work is showing the hardness
of $SPP$. 
 In the univariate case, the hardness of $SPP$ was shown by using a known hard problem called the
Shortest Conjugate Problem ($SCP$). To show the hardness of $SPP$ in the multivariate case we
formulate a new problem called the Smallest Substitution Problem ($SSub$) and show that $SCP$
can be polynomially reduced to $SSub$. In the univariate case,
 $SCP$ is based on the isomorphism of number fields.  
In the multivariate case, the hardness of $SSub$ is
 based on  determining if two functional fields are
isomorphic, which is a known hard problem \citep{pukhlikov:1998:functionfieldisomorphism}. 

\subsection*{Outline of the paper}The rest of the paper is organized as follows. 
In Section~\ref{Preliminaries}, we look at preliminaries relating to lattices and ideal lattices. 
We study cyclic lattices in the multivariate case in Section~\ref{cycliclattices}. 
In Section~\ref{ideallattices}, we prove that only free and finitely generated $\mathbb{Z}$-modules have ideal lattices. 
In Section ~\ref{hardpblms}, we define worst case problems for multivariate ideal lattices and show the hardness of these problems in Section~\ref{hardnessmultivariate}.
In Section ~\ref{hashfunctionsmultivariate}, we show that the hash
functions built from multivariate ideal lattices are collision
resistant. 

\section{Background \& Preliminaries}
\label{Preliminaries}
Let $\Bbbk$ be a field, $A$ a Noetherian commutative ring,
$\mathbb{Q}$ the field of rational numbers, $\mathbb{Z}$ the ring of integers and $\mathbb{N}$ the set of positive
integers including zero. Let $\mathbb{R}^m$ be the $m$-dimensional
Euclidean space.
A polynomial ring in an indeterminate $x$ is denoted by $A[x]$.  $A[x_1,\ldots,x_n]$ denotes the multivariate polynomial ring in indeterminates $x_1,\ldots,x_n$
over $A$.  
A monomial $x_{1}^{\alpha_{1}} \ldots x_{n}^{\alpha_{n}}$ is denoted by   $x^{\alpha}$,
where  $\alpha\in {\mathbb{Z}}_{\ge 0}^n$.
If an ideal $\mathfrak{a}$
in $A[x_1,\ldots,x_n]$ is generated by
polynomials, $f_{1},\ldots, f_{s}$, 
then we write
$\mathfrak{a}= \langle f_{1},\ldots, f_{s} \rangle$. 
We assume that there is a monomial order, $\prec$ on the monomials in $A[x_1,\ldots,x_n]$.
With respect to this monomial order, we have the leading monomial ($\mathrm{lm}$), leading 
coefficient ($\mathrm{lc}$) and leading term ($\mathrm{lt}$) of a polynomial
where $\mathrm{lt}(f) = \mathrm{lc}(f)\mathrm{lm}(f)$.

The set of all integral combinations of $n$ linearly independent vectors $b_1,
\ldots, b_n$ in $\mathbb{R}^m$ $(m\ge n)$ is called a lattice, which is
denoted by $\mathcal{L}(b_1,\ldots,b_n)$. That is,  
$\mathcal{L}(b_1,\ldots,b_n) = \{\sum_{i=1}^{n} x_ib_i \mid x_i \in \mathbb{Z}\}$.
The integers $n$ and $m$ are called the rank and dimension of the
lattice, respectively. The sequence of vectors $b_1, \ldots, b_n$ is
called a lattice basis. When $n=m$, we say that $\mathcal{L}$ is full
rank or full dimensional. An example of $n$-dimensional lattice is the set $\mathbb{Z}^n$ of 
all vectors with integral coordinates. In sequel, whenever we mention
lattices we mean integer lattices,  lattices where the basis vectors
have integer coordinates.   
Integer lattices are additives subgroups of $\mathbb{Z}^N$, $N \in
\mathbb{N}$. 

Determining 
the minimum distance ($\lambda_1$), successive minima ( $\lambda_1, \ldots, \lambda_n$)  and covering radius ($\rho$) of a
lattice, efficiently, are well-known hard problems.  
The approximate algorithms that run in polynomial time give rise to approximation factors that are exponential in the dimension of the lattice. 
In fact,  cryptographic functions based on lattices are built under the assumption that there exists no efficient algorithm that can achieve 
 polynomial approximation factors at most $\gamma(n) = n^{O(1)}$, at least, in the worst case. 
For a
good exposition on lattices and lattice problems one can refer to
\citep{Micciancio:2002:LatticesCrypto}.  

We give below a formal definition of ideal lattices in one variable.
\begin{definition}
Given a monic polynomial  $f\in \mathbb{Z}[x]$ of degree $N$,    
an ideal lattice is an integer lattice $\mathcal{L} \subseteq
\mathbb{Z}^N$ such that it is isomorphic, as a
$\mathbb{Z}$-module, to an ideal, $\mathfrak{A}$ in $\mathbb{Z}[x]/
\langle f \rangle$. 
\end{definition}
The following $\mathbb{Z}$-module homomorphism between $\mathbb{Z}[x]/\langle f \rangle$ and $\mathbb{Z}^N$, where $f$ is a monic polynomial of degree $N$, further elucidates the definition of ideal lattices.
\begin{align*} 
 \psi :  \mathbb{Z}[x]/\langle f \rangle &\longrightarrow \mathbb{Z}^N \\
         \sum\limits_{i=0}^{N-1} a_i x^{i} + \langle f \rangle &\longmapsto (a_0 , \ldots, a_{N-1}).
\end{align*}
Clearly, $\psi$ is a $\mathbb{Z}$-module isomorphism that implies all
$\mathbb{Z}$-submodules (including ideals) in $\mathbb{Z}[x]/\langle f
\rangle$ are isomorphic to $\mathbb{Z}$- submodules of $\mathbb{Z}^N$. Note that $\mathbb{Z}$-submodules of $\mathbb{Z}^N$ are subgroups of 
$\mathbb{Z}^N$ and hence are integer lattices. 
Therefore,  all ideals in $\mathbb{Z}[x]/\langle f \rangle$ are ideal
lattices.

Hash functions are keyed functions that take long strings as inputs and output  short digests that have the following property: 
it is computationally hard to find two distinct 
inputs $x \neq y$ such that $f(x) = f(y)$, where $f$ is a hash
function. Consider the residue class ring,  $\mathbb{Z}_p[x]/\langle f \rangle$, where 
$f \in \mathbb{Z}_p[x]$ is a monic, irreducible polynomial of degree $n$ and $p$ is an integer of order approximately $n^2$. 
A hash
function, $h$, can be designed for ideal lattices in $\mathbb{Z}_p[x]/\langle f \rangle$  by selecting $m$ random elements 
$a_1, \ldots, a_m$ to form an ordered $m$-tuple, $(a_1,\ldots, a_m)$. 
Let $D$ be a strategically chosen subset of $\mathbb{Z}_p[x]/\langle f \rangle$ \citep[Section 5.1]{Lyubashevsky:2006:Ideallatticefirstdef}.
Then the hash function $h$ maps the elements of $D^m$ 
to $\mathbb{Z}_p[x]/\langle f \rangle$ as follows: if $b = (b_1, \ldots, b_m) \in D^m$, then
$h(b) = \sum_{i=1}^m a_i \cdot b_i$. A problem called the ``Shortest
Polynomial Problem'' ($SPP$) equivalent to known hard problems is used 
to prove the collision resistance of the hash function  \citep{Lyubashevsky:2006:Ideallatticefirstdef}. It can be shown that if there
is a polynomial time algorithm that can find a collision with non-negligible probability, then $SPP$ can be solved in polynomial time for every 
lattice in the the ring, $\mathbb{Z}_p[x]/\langle f \rangle$. 

\section{Multivariate Cyclic Lattices}
\label{cycliclattices}
Before we look into the multivariate case we recall the definition of cyclic
lattices. 
\begin{definition}\label{cycliclatticesdef1}
 A lattice $\mathcal{L}$ in $\mathbb{Z}^N$ is a cyclic lattice
if for all $v \in \mathcal{L}$, a cyclic shift of $v$ is also in $\mathcal{L}$.
\end{definition}
One can easily verify the following fact.
\begin{lemma}\label{cycliclatticesdef2}
  A set $\mathcal{L}$ in $\mathbb{Z}^N$ is a cyclic lattice if $\mathcal{L}$ is an ideal in $\mathbb{Z}[x]/\langle x^N-1 \rangle$. 
\end{lemma}
Now consider 
$\mathbb{Z}[x_1,\ldots,x_n]/\langle {x_1}^{r_1} - 1,\cdots, {x_n}^{r_n} - 1\rangle $, 
for some 
$r_{1},\ldots,r_{n} \in \mathbb{N}$. Let $\mathfrak{a} = \langle {x_1}^{r_1} - 1,\cdots, {x_n}^{r_n} - 1\rangle$ and $r_1 \times r_2 \times \cdots \times r_n = N$. 
Then, $\mathbb{Z}[x_1,\ldots,x_n]/\mathfrak{a}$ is a free $\mathbb{Z}$-module, isomorphic to ${\mathbb{Z}}^{N}$ with
$\mathcal{B} = \{{x_1} ^{\alpha_1}\ldots {x_n} 
^{\alpha_n} + \mathfrak{a}, \alpha_k = 0,\ldots,{r_k}-1, k = 1,\ldots,n\}$ as
a $\mathbb{Z}$-module basis. Given an element of the residue class polynomial ring,  
\begin{displaymath} 
\sum_{j=1}^N a_{(\alpha_{1j},\ldots,\alpha_{nj})}{x_1}
 ^{\alpha_{1j}}\ldots {x_n} ^{\alpha_{nj}} + \mathfrak{a},
 \end{displaymath}
 where
 $\alpha_{kj} = 0,\ldots, {r_k}-1$  and
 $a_{(\alpha_{1j},\ldots,\alpha_{nj})} \in \mathbb{Z}$. This can 
 be represented using a tensor, $\mathcal{A} \in
 \mathbb{Z}^{r_1\times\cdots\times r_n}$ defined as
$\mathcal{A}_{i_1, \ldots, i_n} = a_{(i_1 -1, \ldots, i_n - 1)}$,
where ${A}_{i_1, \ldots, i_n}$ denotes $(i_1, \ldots, i_n)$th element in the tensor 
$\mathcal{A}$.  

Now consider $\mathbb{Z}^{N}$ and suppose $r_{1},\ldots,r_{n} \in \mathbb{N}$ such that 
$r_1 \times r_2 \times \cdots \times r_n = N$. Given a lattice 
$\mathcal{L} \subseteq \mathbb{Z}^N$, where 
$ \mathbb{Z}^N = \mathbb{Z} ^{r_1 \times \cdots \times r_n}$, 
it is easy to see that a one-to-one correspondence exists between a vector in 
$\mathcal{L}$ and a tensor in $\mathbb{Z} ^{r_1 \times \cdots \times r_n}$.

Let $\mathcal{A}$ be a tensor in $\mathbb{Z} ^{r_1 \times \cdots \times r_n} $.  
We  define a $(n-1) ^\text{\tiny{th}}$ order tensor  for each 
$i = 1,\ldots,n$ and denote it as $A_{i}(j)$, where 
$A_{i}(j) \in \mathbb{Z} ^{r_1 \times r_2 \times \cdots \times r_{i-1} \times r_{i+1} \times \cdots \times r_n}$, $j = 0,\ldots, r_i - 1$.  
We have,
\begin{displaymath}
A_{i}(j)_{(k_1,\ldots,k_{i-1},k_{i+1},\ldots,k_n)} = \mathcal{A} _{(k_1,\ldots,k_{i-1},j,k_{i+1},\ldots,k_n)}, \hspace{5pt} j = 0,\ldots, r_i - 1.
\end{displaymath} 
We construct the following ordered set of $(n-1)^ \text{\tiny{th}}$ order tensors for each $i = 1,\ldots,n$,
\begin{displaymath}
 \mathcal{A}_{i} = (A_{i}(0), A_{i}(1), \cdots, A_{i}(r_i - 1)).
\end{displaymath}
Using this set, we introduce the notion of multivariate cyclic shifts. 
\begin{definition}
Let $\mathcal{L} \subseteq \mathbb{Z}^N = \mathbb{Z} ^{r_1 \times \cdots \times r_n}$ be a lattice and $\mathcal{A} \in  \mathbb{Z} ^{r_1 \times \cdots \times r_n}$, 
a tensor in $\mathcal{L}$. The ${i}^{\mathrm{th}}$-multivariate cyclic shift of $\mathcal{A}$, $\sigma_{i}(\mathcal{A})$ is a cyclic shift of elements in the ordered set $\mathcal{A}_{i} $. 
\end{definition}
Observe that multiplying an element in $\mathbb{Z}[x_{1},\ldots,x_{n}]/\langle {x_1}^{r_1} - 1,\cdots, {x_n}^{r_n} - 1\rangle$ with $x_i$ 
results in a cyclic shift in the ordered set, $\mathcal{A}_{i}$, $i = 1,\ldots, n$.  This is
also equivalent to a cyclic permutation in the $n^\text{\tiny{th}}$ order tensor along the $i ^\text{\tiny{th}}$ direction. \\
We now formerly define multivariate cyclic lattices. 
\begin{definition}
 A lattice $\mathcal{L}$ in $\mathbb{Z}^N = \mathbb{Z} ^{r_1 \times \cdots \times r_n}$ is a multivariate cyclic lattice if
for all $v \in \mathcal{L}$, a $i^{\mathrm{th}}$-multivariate cyclic shift of $v$ is also in $\mathcal{L}$ for all $i = 1,\ldots,n$.
\end{definition}
\begin{example} \label{cyclicexample}
Consider the case when $n = 3$ and we have $r_1 = 2$, $r_2 = 2$ and $r_3 = 3$. The residue class ring associated to it is $\mathbb{Z}[x_1,x_2,x_3]/\langle {x_1}^2 - 1, {x_2}^2 -1, {x_3}^3 -1\rangle$. It is isomorphic to the space of $3^\text{\tiny {rd}}$ order tensors, $\mathbb{Z}^{2 \times 2 \times 3} (\cong \mathbb{Z}^{12})$. The following set of monomials form the set of coset representatives for a $\mathbb{Z}$-module basis, 
\vspace{-2pt}
\begin{displaymath}
\{1,x_1,x_2, x_3, {x_3}^2, x_1x_2, x_1x_3, x_1{x_3}^2,x_2x_3,x_2{x_3}^2, x_1x_2x_3,x_1x_2{x_3}^2\}.
\end{displaymath} 
Any element in the residue class ring can be represented as a $3^\text{\tiny {rd}}$ order tensor, $\mathcal{A} \in \mathbb{Z}^{2 \times 2 \times 3}$. Let $a_{x^\alpha}$ be the coefficient of the basis element, $x^\alpha$. We can represent $\mathcal{A}$ as follows,\\
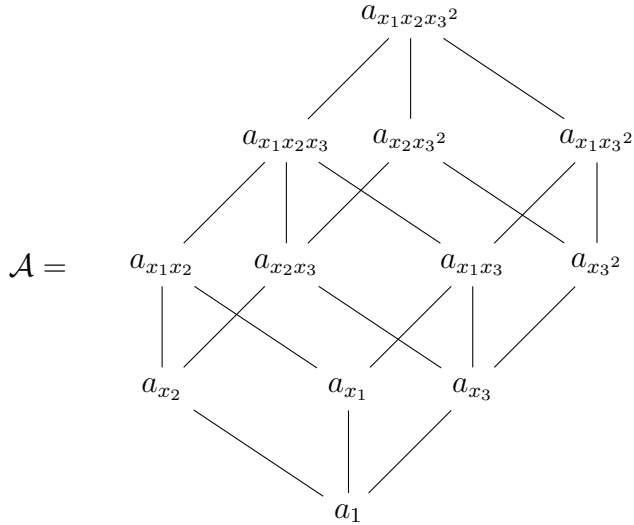
\begin{tikzpicture}[ 
             x = {(4em,4em)} , 
             y ={(-6em, 4em)} ,
             z = {(0em, 4em)}] 
\node (b0) at (-1,1,2) {$\mathcal{A} = $};
\node (b1) at (0,0,0) {$a_1$};
\node (b2) at (1,0,0) {$a_{x_3}$};
\node (b3) at (2,0,0) {$a_{{x_3}^2}$};
\node (b4) at (0,1,0) {$a_{x_2}$};
\node (b5) at (1,1,0) {$a_{{x_2}{x_3}}$};
\node (b6) at (2,1,0) {$a_{{x_2}{x_3}^2}$};
\node (t1) at (0,0,1) {$a_{x_1}$};
\node (t2) at (1,0,1) {$a_{{x_1}{x_3}}$};
\node (t3) at (2,0,1) {$a_{{x_1}{x_3}^2}$};
\node (t4) at (0,1,1) {$a_{{x_1}{x_2}}$};
\node (t5) at (1,1,1) {$a_{{x_1}{x_2}{x_3}}$};
\node (t6) at (2,1,1) {$a_{{x_1}{x_2}{x_3}^2}$};
          
\draw (b1) -- (b2) -- (b3) -- (b6) -- (b5) -- (b4) -- (b1);
\draw (t1) -- (t2) -- (t3) -- (t6) -- (t5) -- (t4) -- (t1);
\draw (t2) -- (t5);
\draw (b2) -- (b5);
 \foreach \x in {1,2,3,4,5,6} {
     \draw (b\x) -- (t\x);
     }
\end{tikzpicture}.\\
The following tensors represent  $A_3(0)$, $A_3(1)$ and $A_3(2)$ respectively. \\
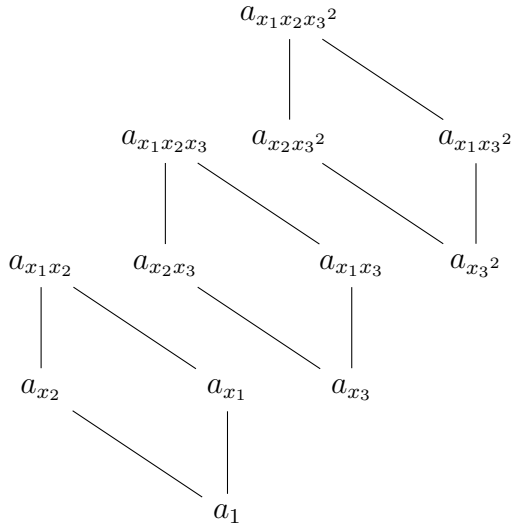
\begin{tikzpicture}[ 
             x = {(4em,4em)} , 
             y ={(-6em, 4em)} ,
             z = {(0em, 4em)}] 
\node (b1) at (0,0,0) {$a_1$};
\node (b2) at (1,0,0) {$a_{x_3}$};
\node (b3) at (2,0,0) {$a_{{x_3}^2}$};
\node (b4) at (0,1,0) {$a_{x_2}$};
\node (b5) at (1,1,0) {$a_{{x_2}{x_3}}$};
\node (b6) at (2,1,0) {$a_{{x_2}{x_3}^2}$};
\node (t1) at (0,0,1) {$a_{x_1}$};
\node (t2) at (1,0,1) {$a_{{x_1}{x_3}}$};
\node (t3) at (2,0,1) {$a_{{x_1}{x_3}^2}$};
\node (t4) at (0,1,1) {$a_{{x_1}{x_2}}$};
\node (t5) at (1,1,1) {$a_{{x_1}{x_2}{x_3}}$};
\node (t6) at (2,1,1) {$a_{{x_1}{x_2}{x_3}^2}$};
          
\draw (b1) --  (b4);
\draw (b2) -- (b5);
\draw (b3) -- (b6);
\draw (t1) --  (t4);
\draw (t2) -- (t5);
\draw (t3) -- (t6);

 \foreach \x in {1,2,3,4,5,6} {
     \draw (b\x) -- (t\x);
     }
\end{tikzpicture}.
%
%
%
\\ $A_3(0)$, $A_3(1)$ and $A_3(2)$ represent $2 ^\text{\tiny{nd}}$ order tensors corresponding to ${x_3}=0$, ${x_3} =1$ and ${x_3}=2$ respectively. Similarly, 
$A_2(0)$ and $A_2(1)$ represent $2^ \text{\tiny{nd}}$ order tensors corresponding to ${x_2}=0$ and ${x_2} =1$ and $A_1(0)$ and $A_1(1)$ represent $2^ \text{\tiny{nd}}$ order tensors corresponding to ${x_1}=0$ and ${x_1} =1$.  Multiplying with ${x_3}$ here results in a cyclic rotation
of $A_3(0)$, $A_3(1)$ and $A_3(2)$. 
\end{example} 
Multiplying with a monomial ${x_1}^{\alpha_1}\cdots{x_n}^{\alpha_n}$ in the general case results in a composition of $\alpha_i$ shifts in  
$\mathcal{A}_{i}$ for each $ i = 1, \ldots, n$. 
The commutativity of multiplication is taken care of as the shifts act on an independent set of subtensors and this makes 
the order of the composition of cyclic shifts irrelevant. That is, the order in which we perform the cyclic shifts between $\mathcal{A}_{i}$ and 
$\mathcal{A}_{j}$ does not matter for 
$i,j = 1,\ldots,n$.  

\begin{proposition}
Every ideal in
\begin{displaymath}
\mathbb{Z}[x_1,\ldots,x_n]/\langle {x_1}^{r_1} - 1,{x_2}^{r_2} - 1,\cdots,
{x_n}^{r_n} - 1\rangle
\end{displaymath}
is a multivariate cyclic lattice.
\end{proposition}

\section{Multivariate Ideal Lattices and Short Reduced Gr\"obner Basis} \label{ideallattices}
 Now we give a formal definition of multivariate ideal lattices. 
\begin{definition}
Given an ideal $\mathfrak{a}\subseteq \mathbb{Z}[x_{1},\ldots,x_{n}]$,    
a multivariate ideal lattice is an integer lattice
$\mathcal{L} \subseteq \mathbb{Z}^N$ that is isomorphic, as a
$\mathbb{Z}$-module, to an ideal
$\mathfrak{A}$ in $\mathbb{Z}[x_{1},\ldots,x_{n}]/ \mathfrak{a}$.  
\end{definition}
In sequel, by ideal lattices we  mean multivariate ideal lattices. 
The $\mathbb{Z}$-module structure of
$\mathbb{Z}[x_{1},\ldots,x_{n}]/\mathfrak{a}$ is crucial  in locating
ideal lattices in  $\mathbb{Z}[x_{1},\ldots,x_{n}]/\mathfrak{a}$. In
general, for a Noetherian ring $A$, one can use Gr\"{o}bner basis
methods to determine an $A$-module representation of
$A[x_{1},\ldots,x_{n}]/\mathfrak{a}$, where $\mathfrak{a}$ is an ideal
in $A[x_{1},\ldots,x_{n}]$
\citep{FrancisDukkipati:2013:freeZmodule}. We describe this briefly
below. 

Consider an ideal $\mathfrak{a} \subseteq A[x_1,\ldots,x_n]$.  Let $G
= \{g_i: i = 1, \ldots, t\}$  be a Gr\"obner basis for $\mathfrak{a}$ w.r.t a monomial order, $\prec$. 
For each monomial, $x^\alpha$, let $J_{x^{\alpha}} = \{i : \mathrm{lm}(g_i)\mid x^{\alpha},  g_i \in G \}$  
and $I_{J_{x^{\alpha}}} = \langle \{\mathrm{lc}(g_i) : i \in
J_{x^{\alpha}}\} \rangle$. 
We refer to $I_{J_{x^{\alpha}}}$ as the leading coefficient ideal
w.r.t. $G$.
Let $C_{J_{x^{\alpha}}}$ represent a set of coset representatives of the equivalence classes in  $A/I_{J_{x^{\alpha}}}$.
Given a polynomial, $f \in  A[x_1,\ldots,x_n]$, let $f =
\sum\limits_{i=1}^m a_i x^{\alpha_i}\hspace{2pt} \mathrm{mod} \hspace{2pt}\langle G \rangle$,
where $a_i \in A, i=1,\ldots,m$.   
If $A[x_{1},\ldots,x_{n}]/\langle G \rangle$ is a finitely generated $A$-module of size $m$,
then corresponding to coset representatives, $C_{J_{x^{\alpha_1}}}, \ldots, C_{J_{x^{\alpha_m}}}$, 
there exists an $A$-module isomorphism, 
\begin{equation} \label{equation}
\begin{split}
 \phi :  A[x_{1},\ldots,x_{n}]/\langle G \rangle &\longrightarrow A/I_{J_{x^{\alpha_1}}} \times \cdots \times A/I_{J_{x^{\alpha_m}}}\\
         \sum\limits_{i=1}^m a_i x^{\alpha_i} + \langle G \rangle &\longmapsto (c_1 +I_{J_{x^{\alpha_1}}}  , \cdots, c_m + I_{J_{x^{\alpha_m}}}) ,
\end{split}
\end{equation}
where $c_i = a_i \text{  mod  } I_{J_{x^{\alpha_i}}}$ and  $c_i \in C_{J_{x^{\alpha_i}}}$. 
We refer to $A/I_{J_{x^{\alpha_1}}} \times \cdots \times A/I_{J_{x^{\alpha_m}}}$ as the $A$-module representation of 
$A[x_{1},\ldots,x_{n}]/\mathfrak{a}$ w.r.t. $G$  (or equivalently w.r.t. $\prec$).  
If $I_{J_{x^{\alpha_i}}} = \{0\} $, we have $C_{J_{x^{\alpha_i}}} = A$,
$\text{for all } i = 1,\ldots,m$.  
This implies  $A[x_{1},\ldots,x_{n}]/\mathfrak{a} \cong A^m$, i.e. $A[x_{1},\ldots,x_{n}]/\mathfrak{a}$ has an $A$-module basis and it is free. 
We say that $A[x_{1},\ldots,x_{n}]/\mathfrak{a} $ has a free $A$-module representation w.r.t. $G$ (or equivalently w.r.t. $\prec$). When $A=\mathbb{Z}$ and 
$\mathbb{Z}[x_{1},\ldots,x_{n}]/\mathfrak{a} \cong \mathbb{Z}^m$, corresponding to every
ideal, $\mathfrak{A}$ in $\mathbb{Z}[x_{1},\ldots,x_{n}]/\mathfrak{a}$, there exists a
subgroup in $\mathbb{Z}^{m}$. Hence the ideals in $\mathbb{Z}[x_{1},\ldots,x_{n}]/\mathfrak{a}$ are indeed ideal lattices. 

To find the various $\mathbb{Z}$-module representations of 
$\mathbb{Z}[x_{1},\ldots,x_{n}]/\mathfrak{a}$, one needs the notion of `short reduced Gr\"obner bases' ~\citep{FrancisDukkipati:2013:freeZmodule}. We describe
this here for  polynomial rings over any Noetherian, commutative ring, $A$.
\begin{definition}
 Let $\mathfrak{a} \subseteq A[x_1,\ldots,x_n]$ be an ideal.  
 A reduced Gr\"obner basis $G$ of $\mathfrak{a}$  is called a short reduced Gr\"obner basis if
 for each $x^\alpha \in \mathrm{lm}(G)$, the length of the generating set of its leading coefficient ideal, $I_{J_{x^{\alpha}}}$ in \eqref{equation}, 
 is minimal. 
\end{definition}
The reduced Gr\"obner basis in the above definition is as described in
\citep{Pauer:2007:Grobnerbasisrings}. 
When  $A = \mathbb{Z}$ in the above definition, 
 the short reduced Gr\"obner basis is the reduced Gr\"obner basis of  $\mathfrak{a}$, 
where the generator of the leading coefficient ideal is taken as the gcd of all generators.
The short reduced Gr\"obner basis is unique for a particular  monomial order and hence once we fix a monomial order, 
$A[x_{1},\ldots,x_{n}]/\mathfrak{a}$ has a unique $A$-module representation. 

\begin{proposition}\label{Proposition for characterization}
Let $\mathfrak{a} \subseteq A[x_1,\ldots,x_n]$ be a non-zero ideal such that $A[x_1,\ldots,x_n]/\mathfrak{a}$ is finitely generated. 
Let $G$ be a  short reduced Gr\"obner basis for $\mathfrak{a}$ w.r.t. some monomial ordering, $\prec$. Then,
$A[x_1,\ldots,x_n]/\mathfrak{a}$ has a free $A$-module representation w.r.t. $\prec$
if and only if $G$ is monic.
\end{proposition}
A monic basis is a basis where the leading coefficients of all its elements are equal to $1$. We have, therefore, the following result for the case when $A = \mathbb{Z}$. 
\begin{theorem}
If the short reduced Gr\"obner basis w.r.t. some monomial ordering is monic, then every ideal in the $\mathbb{Z}$-module,  
$\mathbb{Z}[x_1,\ldots,x_n]/\mathfrak{a}$ is an ideal lattice. 
\end{theorem}
We illustrate this by an example. 
\begin{example}
Let  $\mathfrak{a} = \langle 3 x^2, 5x^2,y \rangle$ be an ideal in
$\mathbb{Z}[x,y]$. The short reduced Gr\"obner basis for the ideal w.r.t. lex order $y\prec x$ is
$G= \{x^2,y\}$. Since $G$ is monic, $\mathbb{Z}[x,y]/\mathfrak{a}$ has a free representation and hence the $\mathbb{Z}$-module is
free and isomorphic to $\mathbb{Z}^2$. All ideals in $\mathbb{Z}[x,y]/\mathfrak{a}$ are ideal
lattices. For example, the ideal generated by $6x+\langle x^2,y
\rangle$ is isomorphic to the lattice, $\mathcal{L}([(0,6)])$. Note
  that here $\mathcal{L}([(0,6)])$ denotes the subgroup generated by
    $(0,6)$ in ${\mathbb{Z}}^{2}$.
\end{example}
%
 Below we show that if $\mathbb{Z}[x_{1},\ldots,x_{n}]/\mathfrak{a}$ is not a free $\mathbb{Z}$-module then it does not contain any ideal lattices. 
\begin{proposition}
If  a finitely generated $\mathbb{Z}$-module,
 $\mathbb{Z}[x_1,\ldots,x_n]/\mathfrak{a}$ is not free then no ideal in
 $\mathbb{Z}[x_1,\ldots,x_n]/\mathfrak{a}$ is an integer lattice. 
 \end{proposition}
 \proof
 We have the following structure theorem over a principal ideal domain (PID),
 \begin{displaymath}
  \mathbb{Z}[x_{1},\ldots,x_{n}]/\mathfrak{a} \cong \mathbb{Z}^l \oplus \mathbb{Z}/\langle w_1 \rangle \oplus \cdots \oplus \mathbb{Z}/\langle w_k \rangle.
 \end{displaymath}
 Clearly, if there is a non zero torsion part in the above direct sum decomposition then $\mathbb{Z}[x_{1},\ldots,x_{n}]/\mathfrak{a}$ will not have a free $\mathbb{Z}$-module 
 representation w.r.t. any Gr\"obner basis. Also, we assume w.l.o.g. that the free part is non zero. 
 Let $G$ be the Gr\"obner basis of the ideal, $\mathfrak{a}$ w.r.t. to some monomial ordering. Consider the isomorphism in \eqref{equation} w.r.t. $G$. 
 Assume there exists an ideal, $\mathfrak{A} \subseteq \mathbb{Z}[x_{1},\ldots,x_{n}]/\mathfrak{a}$ such that it is an integer lattice.  
 Let $x^{\alpha_r} + \mathfrak{a} \in \mathfrak{A}$ be an element such that the leading coefficient ideal of $x^{\alpha_r}$ in $\mathbb{Z}$, $I_{J_{x^{\alpha_r}}}$ 
 is equal to $\{0\}$. 
 This implies that the set of coset representatives, $C_{J_{x^{\alpha_r}}} = \mathbb{Z}$, and therefore the monomial corresponds to the free part in ~\eqref{equation}.
Consider the ideal generated by $x^{\alpha_r} + \mathfrak{a}$. Since the $\mathbb{Z}$-module is not free we have  $I_{J_{x^{\alpha_j}}} \ne \{0\}$ and $C_{J_{x^{\alpha_j}}} \ne \mathbb{Z}$
for some monomial $x^{\alpha_j}$ 
in \eqref{equation}. 
Let $c \in C_{J_{x^{\alpha_j}}}$.
Since $c_i x^{\alpha_i} + \mathfrak{a} \in  \mathbb{Z}[x_{1},\ldots,x_{n}]/\mathfrak{a}$, 
 $c x^{\alpha_j }x^{\alpha_r}  + \mathfrak{a} \in \langle x^{\alpha_r} + \mathfrak{a} \rangle $. This implies, the ideal generated by a free element contains torsion elements. 
 Thus the $\mathbb{Z}$-module,  
 $\mathfrak{A}$ has torsion elements and is not isomorphic to an integer lattice, which is a contradiction.
 \endproof
 \begin{corollary}
  Every ideal, $\mathfrak{a}$ in  $\mathbb{Z}[x_{1},\ldots,x_{n}]$ is
  an ideal lattice if and only if
  $\mathbb{Z}[x_{1},\ldots,x_{n}]/\mathfrak{a}$ is a free and finitely
  generated $\mathbb{Z}$-module. 
 \end{corollary}

We recall that in the definition of ideal lattices in $\mathbb{Z}[x]$  the
choice of the polynomial $f$ in $\mathbb{Z}[x]/\langle f \rangle$ is
restricted to monic polynomials.  But in the construction of many
cryptographic primitives like collision resistant hash functions $f$ is assumed to be an irreducible polynomial. 
This condition ensures that the ideal lattice is full rank and hence prevents easy collision attacks
\citep{Lyubashevsky:2006:Ideallatticefirstdef}.
In the multivariate case, we derive a necessary and sufficient  condition
for full rank ideal lattices.  
\begin{proposition}
 Let $\{g_1,\ldots,g_t\}$ be a monic short reduced Gr\"obner basis of an ideal $\mathfrak{a}$ in $\mathbb{Z}[x_1,\ldots,x_n]$  
 such that 
 $\mathbb{Z}[x_1,\ldots,x_n]/\mathfrak{a} \cong \mathbb{Z}^N$ for some $N \in \mathbb{N}$.
All ideals in $\mathbb{Z}[x_1,\ldots,x_n]/\mathfrak{a}$ are full rank
lattices if and only if $\mathfrak{a}$ is a prime ideal.
\end{proposition}
\proof
Let $\mathfrak{a} = \langle g_1,\ldots,g_t \rangle$ be a prime ideal. Consider an ideal $\mathfrak{A} = \langle f_1 + \mathfrak{a}, \ldots,  f_s + \mathfrak{a} \rangle$ in 
$\mathbb{Z}[x_1,\ldots,x_n]/\mathfrak{a}$, where $f_1, \ldots, f_s \in \mathbb{Z}[x_1, \ldots, x_n]$.
Since $\mathbb{Z}[x_1,\ldots,x_n]/\mathfrak{a} \cong \mathbb{Z}^N$ we have a finite basis, $\mathcal{B} = \{b_1 + \mathfrak{a}, \ldots,b_N + \mathfrak{a}\}$.
We have to prove that there are $N$ linearly independent vectors in $\mathfrak{A}$.
Consider $f_1b_1,\ldots,f_1b_N$. Let $c_1f_1b_1 + \cdots + c_Nf_1b_N \in \langle g_1, \ldots, g_t\rangle$. This implies $f_1(c_1b_1 + \cdots + c_Nb_N) \in \langle g_1, \ldots, g_t\rangle$. 
Since $\langle g_1,\ldots,g_t \rangle$ is a prime ideal, either
$f_1 \in \langle g_1,\ldots,g_t \rangle$ or $(c_1b_1 + \cdots + c_Nb_N)\in\langle g_1,\ldots,g_t \rangle$. But both cases cannot happen. Therefore $c_i = 0$ for all $i = 1, \ldots,N$.
This implies that  $f_1b_1 + \mathfrak{a},\ldots,f_1b_N + \mathfrak{a}$ are linearly independent and the ideal lattice is full rank.

Conversely, assume that $\mathfrak{a}$ is not a prime ideal. 
Then there exists $l,h \in  \mathbb{Z}[x_1,\ldots,x_n]$ such that $lh \in \langle g_1,\ldots,g_t \rangle$ but $l \not\in \langle g_1,\ldots,g_t \rangle$ and 
$h \not\in \langle g_1,\ldots,g_t \rangle$. This implies, $l = \sum_{i=1}^{N} c_ib_i$ and $h = \sum_{i=1}^{N} d_ib_i$, where $b_i + \mathfrak{a} \in \mathcal{B}$, the basis
for $\mathbb{Z}[x_1,\ldots,x_n]/\mathfrak{a}$ and $c_i,d_i \in \mathbb{Z}$. 
Consider the ideal lattice $\langle l + \mathfrak{a}\rangle$. We have $lh \in \langle g_1,\ldots,g_t\rangle$ and this implies $l\sum_{i=1}^{N}d_ib_i \in \langle g_1,\ldots,g_t\rangle$. 
But $l\not\in \langle g_1,\ldots,g_t\rangle$ and $\sum_{i=1}^{N} d_ib_i \not\in \langle g_1,\ldots,g_t\rangle$. The set $\{lb_1 + \mathfrak{a}, \ldots, lb_N + \mathfrak{a}\}$ 
contains linearly dependent vectors and the rank 
of the ideal lattice $\langle l + \mathfrak{a}\rangle$ is $\lneq N$. 
Therefore, if the ideal $\mathfrak{a}$ is not a prime ideal then there exist lattices in $\mathbb{Z}[x_1,\ldots,x_n]/\mathfrak{a}$ that are not full rank.
\endproof
Determining if an ideal is prime or not is important for many practical applications.
An algorithm for primality testing in polynomial rings, over any commutative, Noetherian ring, $A$ can be found in \citep{Gianni:1988:primailitytesting}. 

We now give an example of a class of binomial ideals that is prime and gives rise to free residue class polynomial rings. 
Given an integer lattice, $\mathcal{L}$, a lattice ideal, $\mathfrak{a}_\mathcal{L}$ in $\Bbbk[x_1,\ldots,x_n]$ is defined as the binomial ideal generated by $\{x^{v^+} - x^{v^-}\}$ 
where $v^+$ and $v^-$ are non-negative with disjoint support and $v^+ - v^- \in \mathcal{L}$ \citep{Katsabekis:2010:latticeideal}. 
Lattice ideals in polynomial rings over $\mathbb{Z}$ can be defined in the same way.  
In this case, the binomial ideal is generated over the polynomial ring, $\mathbb{Z}[x_1,\ldots,x_n]$.  
The generators of the ideal are binomials with the terms having opposite sign and the coefficients of both the terms equal to absolute value $1$. 
One can show that the short reduced Gr\"obner basis of the lattice ideal is monic \citep{FrancisDukkipati:2013:freeZmodule}. 
In this case, by Proposition ~\ref{Proposition for characterization}, $\mathbb{Z}[x_1,\ldots,x_n]/\mathfrak{a}_\mathcal{L}$ is free. 
Hence, we have the following fact. 
\begin{theorem}
Every ideal in $\mathbb{Z}[x_1,\ldots,x_n]/\mathfrak{a}_\mathcal{L}$,
where $\mathfrak{a}_\mathcal{L}$ is a lattice ideal,  is an ideal lattice. 
\end{theorem}
\noindent The saturation of an integer lattice, $\mathcal{L}\subseteq \mathbb{Z}^m$ is a lattice, defined as  
 \begin{displaymath}
  Sat(\mathcal{L}) = \{\alpha \in \mathbb{Z}^m \mid d\alpha \in \mathcal{L} \text{ for some } d \in \mathbb{Z}, d \ne 0\}.
 \end{displaymath}
 We say that an integer lattice $\mathcal{L}$ is saturated if $\mathcal{L} =  Sat(\mathcal{L})$. It can be easily shown that the lattice ideal $\mathfrak{a}_\mathcal{L}$ is prime if and only if
$\mathcal{L}$ is saturated. Note that in the commutative algebra literature  prime lattice ideals are also called toric ideals \citep{Bigatti:1999:toricideals}.
Thus, toric ideals in $\mathbb{Z}[x_1,\ldots,x_n]$  give rise to full rank integer lattices.
\section{Hard Problems for Multivariate Ideal Lattices}
\label{hardpblms}
\subsection{Expansion Factor}
\label{expansionfactormultivariate}
\noindent
Given $f \in \mathbb{Z}[x_1, \ldots, x_n]$, the following norms can be defined on $\mathbb{Z}[x_1, \ldots, x_n]$: the infinity norm ${\lVert f \rVert}_\infty$  
that takes the maximum coefficient of all the terms in the polynomial 
and  the norm w.r.t. an ideal $\mathfrak{a}$ and a monomial order  $\prec$, ${\lVert f
  \rVert}_{\mathfrak{a},\prec}$ that takes the maximum coefficient of all 
the terms in the polynomial reduced modulo $\mathfrak{a}$ w.r.t. $\prec$.

Given a finitely generated residue class polynomial ring 
$\mathbb{Z}[x_1,\ldots,x_n]/\mathfrak{a}$ with a free $\mathbb{Z}$-module representation w.r.t a monomial order $\prec$, the ideal $\mathfrak{a}$
should satisfy the following properties that are essential for the
security proofs of the hash function: (i)  $\mathfrak{a}$
should be a prime ideal, which ensures that every ideal in
$\mathbb{Z}[x_1,\ldots,x_n]/\mathfrak{a}$ is a full rank lattice, and (ii) the norm of any
polynomial $f$ w.r.t. the ideal $\mathfrak{a}$ and monomial order $\prec$, 
${\lVert f \rVert}_{\mathfrak{a},\prec}$ should not be much larger than   
${\lVert f \rVert}_\infty$. 
The second property is formally captured with a parameter called the
expansion factor that we define for the multivariate case below.  

For a given finite set of generators, $\mathrm{maxdeg}_{x_i}(\mathfrak{a})$ denotes the maximum
degree of a variable $x_i$ among the generators of the ideal
$\mathfrak{a}$. We represent the
maximum degree of a variable $x_i$ in a polynomial $g$ as $\mathrm{maxdeg}_{x_i}(g)$.   
\begin{definition}
 Let $\mathfrak{a} = \langle f_1, \ldots, f_s \rangle \subseteq \mathbb{Z}[x_1,\ldots,x_n]$ such that $\mathbb{Z}[x_1,\ldots,x_n]/\mathfrak{a}$ is finitely generated 
 and has a free $\mathbb{Z}$-module representation w.r.t. $\prec$. The expansion factor $\mathcal{E}$ of $\mathfrak{a}$ is defined as
 \begin{displaymath}
  \mathcal{E}(\mathfrak{a}, \prec, (k_1, \ldots, k_n)) = 
  \max_{\substack{\mathrm{maxdeg}_{x_i}(g) \leq k_i(\mathrm{maxdeg}_{x_i}(\mathfrak{a}))\\ \forall i \in \{1, \ldots, n\} \\ g\in\mathbb{Z}[x_1,\ldots,x_n]}} \frac{{\lVert g \rVert}_{\mathfrak{a}, \prec}}{{\lVert g \rVert}_\infty}, 
 \end{displaymath}
 where $k_i \in \mathbb{N}$, $i = 1,2,\ldots, n$.
 \end{definition}
 We give a result that bounds  the expansion factor of ideals
for which the residue class polynomial ring is  finitely
generated and has a free $\mathbb{Z}$-module representation.  
\begin{theorem}
Let $G =\{g_1, \ldots, g_s\}$ be a short reduced Gr\"obner basis of an ideal
$\mathfrak{a}$  w.r.t. a monomial order $\prec$
such that $\mathbb{Z}[x_1,\ldots,x_n]/\mathfrak{a}$ is finitely
generated and has a free $\mathbb{Z}$-module representation w.r.t. $\prec$ (i.e. $G$ is monic). Then for any $f \in
\mathbb{Z}[x_1,\ldots,x_n]$, ${\lVert f \rVert}_{\mathfrak{a},\prec} \leq {\lVert f \rVert}_\infty {(2\cdot {({\lVert g \rVert}_\infty)}_\mathrm{max})}^k$, where 
${({\lVert g \rVert}_\infty)}_\mathrm{max}$
denotes the maximum norm among the generators of the ideal and $k$ is of the order $O({(\mathrm{deg}(f))}^n{(\max\limits_{1\leq i\leq s}{\mathrm{deg} (g_i)})}^n)$.
\end{theorem} 
\proof
First we reduce $f$ with the generators $\{g_1, \ldots, g_s\}$. Let $g_j$ be the generator such that $\mathrm{lm}(f) = x^\alpha \mathrm{lm}(g_j) $ for 
some $x^\alpha$. Then, $f_1 = f - \mathrm{lc}(f)x^\alpha g_j$. Since
$G$ is monic, during the reduction process one needs to consider only one generator of the ideal at a
time.   
 We have,
 \begin{align*}
 \lVert f_1 \rVert_\infty  \leq  \lVert f\rVert_\infty +  \lVert f \rVert_\infty  \lVert g_j\rVert_\infty
&\leq 2 \lVert f \rVert_\infty   \lVert g_j \rVert_\infty \\
 &\leq 2 \lVert f \rVert_\infty  {({\lVert g \rVert}_\infty)}_\mathrm{max}.
\end{align*}
Next we can reduce $f_1$ by any of the generators in the Gr\"obner basis to get $f_2$ and continue this process. This process will terminate after $k$ steps, 
where $k$ is of the order $O({(\mathrm{deg}(f))}^n{(\max\limits_i{\mathrm{deg} (g_i)})}^n)$ \citep{thieu:2013:reduction}.
The exact number of iterations cannot be determined unless we know the
exact structure of the ideal and the polynomial. Hence,
\begin{displaymath}
{\lVert f \rVert}_{\mathfrak{a},\prec} \leq {\lVert f \rVert}_\infty {(2\cdot {({\lVert g \rVert}_\infty)}_\mathrm{max})}^k.
\end{displaymath}
\endproof
\subsection{Worst Case Problems}
\label{worstcasepblms}
\noindent
For any ideal
$\mathfrak{A} \subseteq \mathbb{Z}[x_1,\ldots,x_n]/\mathfrak{a}$ 
 we use ${\lambda_i}^p(\mathfrak{A})$ to indicate
 ${\lambda_i}^p(\mathcal{L}(\mathfrak{A}))$, where ${\lambda_i}$ represents the $i$-th successive minima w.r.t. the $\ell_p$ norm. 
 \begin{definition}
 The approximate Shortest Polynomial Problem
 ($SPP_\gamma(\mathfrak{A})$) is defined as follows: 
 given an ideal $\mathfrak{A}\subseteq
 \mathbb{Z}[x_1,\ldots,x_n]/\mathfrak{a}$, where
 $\mathbb{Z}[x_1,\ldots,x_n]/\mathfrak{a}$ is finitely
 generated and has a free $\mathbb{Z}$-module representation w.r.t. $\prec$,  
 determine a $g \in\mathfrak{A}$ such that $g \neq 0$ and ${\lVert g
   \rVert}_{\mathfrak{a},\prec} \leq \gamma
 {\lambda_1}^\infty(\mathfrak{A})$, where $\lambda_1$ represents the minimum distance. 
 \end{definition}
 We use the notation $\mathcal{L}(\mathfrak{a})$ to denote the set of all lattices
 associated with $\mathbb{Z}[x_1,\ldots,x_n]/\mathfrak{a}$ and use $\mathfrak{a}-SPP$  when we consider $SPP$ for ideals in  $\mathbb{Z}[x_1,\ldots,x_n]/\mathfrak{a}$, where
 $\mathfrak{a}$ is as described above.
 In Section \ref{hardnessmultivariate}, we show how well known hard problems can be reduced to $\mathfrak{a}-SPP_\gamma$. 
 
 We give below a lemma that relates ${\lambda_1}^\infty$ with ${\lambda_N}^\infty$ for an ideal $\mathfrak{A} \subseteq \mathbb{Z}[x_1,\ldots,x_n]/\mathfrak{a}$, 
 where $\mathfrak{a}$ is a prime ideal and $\mathbb{Z}[x_1,\ldots,x_n]/\mathfrak{a}$ is free and finitely generated of dimension $N$. 
It shows that ${\lambda_N}^\infty$ cannot be much bigger than ${\lambda_1}^\infty$ if the ideal is prime.
\begin{lemma}\label{lemma4.2}
 For every ideal $\mathfrak{A}  \subseteq
 \mathbb{Z}[x_1,\ldots,x_n]/\mathfrak{a}$, where $\mathfrak{a}$ is a
 prime ideal and $\mathbb{Z}[x_1,\ldots,x_n]/\mathfrak{a}$ is
 finitely generated of size $N$ and has a free $\mathbb{Z}$-module representation w.r.t. $\prec$, we have   
\begin{displaymath} 
 {\lambda_N}^\infty(\mathfrak{A})  \leq \mathcal{E}(\mathfrak{a}, \prec, (2,\ldots,2)) {\lambda_1}^\infty(\mathfrak{A}) .
 \end{displaymath}
\end{lemma}
\proof
Let $g$ be a polynomial in $\mathfrak{A}$ reduced w.r.t. $\mathfrak{a}$ such that $\lVert g \rVert _{\infty} = {\lambda_1}^\infty(\mathfrak{A}) $.
Let $\mathcal{B}=\{b_1, \ldots, b_N\}$ be the basis for $\mathbb{Z}[x_1,\ldots,x_n]/\mathfrak{a}$.
Then $\{gb_1, \ldots, gb_N\}$ is a linearly independent set because $\mathfrak{a}$ is a prime ideal.
Also, $\mathrm{maxdeg}_{x_i}(gb_i) \leq
2\cdot\mathrm{maxdeg}_{x_i}(\mathfrak{a})$. For $i = 1, \ldots, N$,
\begin{align*}
  \lVert gb_i \rVert _{\mathfrak{a},\prec} \leq \mathcal{E}(\mathfrak{a},\prec, (2,\ldots,2)) \lVert   gb_i \rVert _{\infty}
  &\leq  \mathcal{E}(\mathfrak{a},\prec, (2,\ldots,2)) \lVert  g \rVert _{\infty},\\
  & = \mathcal{E}(\mathfrak{a},\prec, (2,\ldots,2)){\lambda_1}^\infty(\mathfrak{A}).
\end{align*}
\endproof
Now, we define an incremental version of $SPP$. 
\begin{definition}
The approximate 
Incremental Shortest Polynomial Problem
$(IncSPP_\gamma(\mathfrak{A},g))$ is defined as follows: 
 Given an ideal $\mathfrak{A} \subseteq
 \mathbb{Z}[x_1,\ldots,x_n]/\mathfrak{a}$ and $g\in \mathfrak{A}$ 
such that $\lVert g \rVert_{\mathfrak{a},\prec} \lneq \gamma
{\lambda_1}^\infty(\mathfrak{A})$, determine an $h \in\mathfrak{A}$
such that 
$\lVert h \rVert_{\mathfrak{a},\prec} \neq 0$ and $\lVert h
\rVert_{\mathfrak{a},\prec} \leq \lVert g \rVert_{\mathfrak{a},\prec} /2$.    
\end{definition}
The following result directly follows.
\begin{lemma}
 There is a polynomial time reduction from $\mathfrak{a}-SPP_\gamma$ to $\mathfrak{a}-IncSPP_\gamma$.
\end{lemma}
\section{Hardness Results}
 \label{hardnessmultivariate}
\noindent 
 Let $\mathfrak{a}$ and $\mathfrak{a}^{'}$ be ideals in
$\mathbb{Z}[x_{1},\ldots,x_{n}]$ defined as
$\mathfrak{a} = \langle {x_1}^{r_1} - 1,{x_2}^{r_2} - 1,\ldots,
{x_n}^{r_n} - 1\rangle,  r_i \in \mathbb{N}, i = 1,2, \ldots,n,$ and 
$\mathfrak{a}^{'}  = \langle {x_1}^{r_1 - 1}+ {x_1}^{r_1 -
  2}+\cdots+1, \ldots,{x_n}^{r_n -1}+{x_n}^{r_n -2}+\cdots +
1\rangle$. 
We  prove  that solving  $SPP_\gamma$ in an ideal in
$\mathbb{Z}[x_1,\ldots,x_n]/\mathfrak{a}$ is equivalent to finding the
approximate shortest polynomial in $\mathbb{Z}[x_1,\ldots,x_n]/\mathfrak{a}^{'}$. 
Note that if each $r_i$ is a prime number then $\mathfrak{a}^{'}$
is a prime ideal and we have full rank lattices. It also means that each of the generators is irreducible.
If one can solve the approximate shortest polynomial problem in the ideal lattices of 
$\mathbb{Z}[x_1,\ldots,x_n]/\mathfrak{a}^{'}$,
 then one can also solve the approximate shortest polynomial problem in multivariate  cyclic lattices (where each $r_i$ is prime), that we conjecture is a hard problem. 
\begin{lemma}
Let $\mathfrak{A}$ be an ideal in
$\mathbb{Z}[x_1,\ldots,x_n]/\mathfrak{a}$ such that the residue class
polynomial ring is finitely generated of size $N$ and has a free $\mathbb{Z}$-module representation w.r.t. $\prec$. Given
the generators for $\mathfrak{A}$, there is a polynomial time
algorithm to find the basis for the lattice of $\mathfrak{A}$,
$\mathcal{L}(\mathfrak{A})$.  
\end{lemma}
\proof
Let $\mathfrak{A} = \{ g_{1}+\mathfrak{a},\ldots,g_{m}+\mathfrak{a}\}$.
Let the residue classes of $\mathcal{B} = \{b_1,
\ldots, b_N \}$ be a basis for
$\mathbb{Z}[x_1,\ldots,x_n]/\mathfrak{a}$.  Consider the set $ G = \{
g_1b_1 + \mathfrak{a}, \ldots, g_1b_N + \mathfrak{a}, \ldots, g_mb_1
+ \mathfrak{a}, \ldots, g_mb_N + \mathfrak{a}\}$. 
All the elements of $\mathfrak{A}$ can be written as an integer combination of elements in $G$ and therefore 
$\mathfrak{A}$ is a $\mathbb{Z}$-module. Using Hermite normal form one can determine the basis of the $\mathbb{Z}$-module as an additive group  in polynomial time. 
\endproof
\begin{lemma}
Let $\mathfrak{a}$ and $\mathfrak{a}^{'}$ be ideals as defined as above.
Given a multivariate cyclic lattice $\mathfrak{A}$ in  
$\mathbb{Z}[x_1,\ldots,x_n]/\mathfrak{a}$ of dimension $N$, there is a polynomial time reduction from the problem of approximating the shortest vector in 
$\mathfrak{A}$ within a factor of $2\gamma$ to approximating the shortest vector in an ideal in the ring, 
$\mathbb{Z}[x_1,\ldots,x_n]/\mathfrak{a}^{'}$
 within a factor of $\gamma$. 
\end{lemma}
\proof
Let  $f$ be a polynomial of smallest infinity norm  such that $f + \mathfrak{a} \in  \mathfrak{A}$ and  $f + \mathfrak{a}$ is 
reduced modulo $\mathfrak{a}$ w.r.t. some monomial order, $\prec$. 
If $f \notin  {\mathfrak{a}}^{'}$, ${\lVert f
  \rVert}_{\mathfrak{a}^{'},\prec} \leq 2 {\lVert f \rVert}_{\infty}$, since its residue class is
 reduced w.r.t.  $\mathfrak{a}^{'}$. There
exists a non zero polynomial in $\mathfrak{A}$ whose infinity norm is
at most  $ 2 {\lVert f \rVert}_{\infty}$. Thus the algorithm for
approximating the shortest polynomial 
in  $\mathbb{Z}[x_1,\ldots,x_n]/\mathfrak{a}^{'}$ to within a factor of  $\gamma$
will  find a non-zero polynomial of infinity norm at most $2\gamma
{\lVert f \rVert}_{\infty}$. 
Every non-zero polynomial in
$\mathbb{Z}[x_1,\ldots,x_n]/\mathfrak{a}^{'}$ is non zero in
$\mathbb{Z}[x_1,\ldots,x_n]/\mathfrak{a}$. 
If $f \in  \mathfrak{a}^{'}$, 
we have $f \in {\mathfrak{a}^{'}\cap \mathfrak{A}}$.  
Since $f$ is  reduced w.r.t. $\mathfrak{a}$, $f$ is a sum of integer multiples of the generators of $\mathfrak{a}^{'}$.
We can find a basis for the one dimensional lattice ${\mathfrak{a}^{'}\cap \mathfrak{A}}$ and the generator will be the shortest polynomial.

\begin{conjecture}
Approximation problems like $SVP_\gamma$ are computationally hard in
multivariate cyclic lattices with prime powers.   
\end{conjecture}
The conjecture is based on the assumption that the $SVP_\gamma$ problem is hard for univariate cyclic lattices of prime powers \citep{Micciancio:2002:CyclicLattices}. Given, 
\begin{displaymath}
\mathbb{Z}[x_1,\ldots,x_n]/\langle {x_1}^{r_1} - 1,{x_2}^{r_2} - 1,\cdots,
{x_n}^{r_n} - 1\rangle, \hspace{5pt} r_i \in \mathbb{N}, 
\end{displaymath}
where each $r_i$   is prime, the multivariate cyclic lattice in $n$ indeterminates is  equivalent to $n$ independent  univariate cyclic lattices  of prime powers.  
This is because the multivariate cyclic shifts in the $n^ \text{\tiny{th}}$ order tensor $\mathcal{A}_i$ for each $i = 1,\ldots,n$ are independent of each other (see Section~\ref{cycliclattices}). 
This implies, the assumption that the $SVP_\gamma$ problem is hard for univariate cyclic lattices of prime powers  can be applied for each $i = 1,2,\ldots,n$  individually. 
Therefore, if the approximation problems are hard for univariate
cyclic lattices with prime powers then they are computationally hard
for multivariate cyclic lattices with prime powers as well.

We now give the hardness results for multivariate  ideal lattices 
based on results from function fields of algebraic varieties.
A function field of an affine variety $\mathcal{V}$ is the quotient field of the coordinate ring $\Bbbk[x_1, \ldots,x_n]/\mathcal{I}(\mathcal{V})$, 
often described as the  field of rational functions on $\mathcal{V}$.
Note that in the univariate case, the $SPP$ problem can be reduced to
the problem of finding small conjugates in ideals of subrings of a
number field which is a hard problem~\citep{Lyubashevsky:2006:Ideallatticefirstdef}.  

To prove the hardness of $SPP$ we define the following problem.
Let $\mathfrak{a}$ be an ideal in $\mathbb{Z}[x_1,\ldots,x_n]$ such
that $\mathbb{Z}[x_1,\ldots,x_n]/\mathfrak{a}$ is free and finitely
generated. Consider the variety of $\mathfrak{a}$ in $\mathbb{C}^n$, $\mathcal{V}_{\mathbb{C}}(\mathfrak{a})$. 
Then 
for every $(a_1, \ldots, a_n) \in \mathcal{V}_{\mathbb{C}}(\mathfrak{a})$ 
the following mapping 
 \begin{equation}\label{isomorphism}
\begin{aligned}  
   \psi : \mathbb{Z}[a_1,\ldots,a_n] &\longrightarrow \mathbb{Z}[x_1,\ldots,x_n]/\sqrt{\mathfrak{a}}
  \\ \sum_{i=1}^{l}{\alpha_i {a_1}^{i_1}\ldots {a_n}^{i_n}} &\longmapsto \sum_{i=1}^{l}\alpha_i {{x_1}^{i_1}\ldots {x_n}^{i_n} + \sqrt{\mathfrak{a}}},
 \end{aligned}
 \end{equation}
 where $l \in \mathbb{N}$ and $\sqrt{\mathfrak{a}}$ is the radical of the ideal, is an isomorphism. 
 When $\mathbb{Z}[x_1,\ldots,x_n]/\sqrt{\mathfrak{a}}$ is free and finitely generated, $\mathcal{V}_{\mathbb{C}}(\mathfrak{a})$ is a finite set. For ease of notation we will omit the subscript $\mathbb{C}$ and denote the variety as $\mathcal{V}(\mathfrak{a})$.

For $(a_1, \ldots, a_n)$ $\in \mathcal{V}(\mathfrak{a})$ and
$\alpha = \sum_{i=1}^{l}{\alpha_i{a_1}^{i_1}\cdots {a_n}^{i_n}}$, a
polynomial in $\mathbb{Z}[a_1,\ldots,a_n]$,  we define
$\mathrm{maxCoeff}_{(a_1, \ldots, a_n)}(\alpha)$ as
$\max\limits_{1\leq i \leq l}(\mid \alpha_i \mid)$. 
Let $\psi_j$ be the
isomorphism defined as in Equation~\eqref{isomorphism} for each element of the affine variety, $\mathcal{V}(\mathfrak{a})$.  
Given an ideal $I$ in $\mathbb{Z}[a_1, \ldots, a_n]$, $(a_1, \ldots, a_n) \in \mathcal{V}(\mathfrak{a})$, for an
element $\alpha =\sum_{i=1}^{l}{\alpha_i {{a_1}}^{i_1}\ldots
  {{a_n}}^{i_n}} $ in  $I$, we define 
 \begin{displaymath}
  \mathrm{maxsub}(\alpha) = 
 \max\limits_{1\leq j \leq N}\{\sum_{i=1}^{l}{\alpha_i
     {{a_1}^{(j)}}^{i_1} \ldots {{a_n}^{(j)}}^{i_n}}: ({a_1}^{(j)},
   \ldots, {a_n}^{(j)}) \in \mathcal{V}(\mathfrak{a}) \}. 
 \end{displaymath}
\begin{definition}(Approximate Smallest Substitution Problem ($SSub$))
 Let $\mathfrak{a} \subseteq \mathbb{Z}[x_1,\ldots,x_n]$ be an ideal such that  $\mathbb{Z}[x_1,\ldots,x_n]/\mathfrak{a} $ 
 is free and finitely generated. Let the finite variety, $\mathcal{V}(\mathfrak{a})$ be of cardinality $N$. Given an ideal $I$ in 
 $\mathbb{Z}[a_1, \ldots, a_n]$, $(a_1, \ldots, a_n) \in \mathcal{V}(\mathfrak{a})$,
 the approximate smallest substitution problem, ${SSub}_\gamma(I)$ is defined as follows:  find an element $\alpha \in I$ such that $\mathrm{maxsub}(\alpha) \leq \gamma \mathrm{maxsub}(\alpha^{'})$, for all
 $\alpha^{'} \in I$. 
\end{definition}

It is important to note that formulation of the smallest substitution problem in  the multivariate case is quite different from the univariate case. In the univariate case, the problem that is
mapped to $SPP$ is the smallest conjugate problem ($SCP$). For any $\alpha$ in the ideal $I$, first a function called $\mathrm{maxConj}$ analogous to the
$\mathrm{maxsub}$ is defined. The function returns the maximum of the zeroes of the minimum polynomial of $\alpha$ over $\mathbb{Q}$. $SCP$ poses the problem of finding an $\alpha \in I$ 
such that it has the least $\mathrm{maxConj}$ among all the elements in $I$. 
This relates to the problem of isomorphism of number fields for which no polynomial time algorithm is 
determined \citep[Polynomial Reduction Algorithm]{cohen:2013:course}. The hardness of $SCP$ is discussed in 
\citep{Lyubashevsky:2006:Ideallatticefirstdef}. We argue that the smallest substitution problem, $SSub$, relates to the problem of isomorphism of function fields, 
the multivariate extension of number fields and a hard problem \citep{pukhlikov:1998:functionfieldisomorphism}. 
We show below that $SCP$ is a special instance of the $SSub$ problem. That is,  $SCP$ is polynomially reducible to $SSub$. 
\begin{theorem}\label{hardnessforspecificideal}
Given an monic irreducible polynomial $f \in \mathbb{Z}[x]$ of degree $N$, let $\mathfrak{a} = \langle f \rangle$ be an ideal in $\mathbb{Z}[x]$.
There is a polynomial time reduction from $\mathfrak{a}-SCP$ to $\mathfrak{a}-SSub$. 
\end{theorem}
\proof
Let $\mathcal{V}$ be the variety associated with $\mathfrak{a}$ of cardinality $N$. For $a \in \mathcal{V}$, we have the isomorphism, $\psi$ given by Equation~\eqref{isomorphism},
  $\mathbb{Z}[a] \cong \mathbb{Z}[x]/\mathfrak{a}$. An algorithm for $\mathfrak{a}-SSub_\gamma$ returns an $\alpha \in  \mathbb{Z}[a]$, $a \in \mathcal{V}(\mathfrak{a})$ 
  such that 
  $\mathrm{maxsub}(\alpha) \leq \gamma \mathrm{maxsub}(\alpha^{'})$, for all
 $\alpha^{'} \in \mathbb{Z}[a]$.  Let $\alpha = \alpha_0 + \alpha_1 a + \cdots + \alpha_{N-1} a^{N-1}$ and 
 \begin{displaymath}
   \mathrm{maxsub}(\alpha) = 
 \max\limits_{1\leq j \leq N}\{\sum_{i=0}^{N-1}{\alpha_i
     {a^{(j)}}^{i}} : a^{(j)} \in \mathcal{V}(\mathfrak{a}) \}. 
 \end{displaymath}
Since the set, $\{\sum_{i=0}^{N-1}{\alpha_i {a^{(j)}}^{i}} \}$ is the set of zeroes of the minimal polynomial 
of $\alpha$ over $\mathbb{Q}$, we have $\mathrm{maxsub}(\alpha) = \mathrm{maxConj}(\alpha)$. 
     Therefore, $\alpha$ is the solution for $\mathfrak{a}-SCP_\gamma$ as well.   
\endproof

We proceed to find a relation between the maximum coefficient of an element $\alpha$ in the ideal $\mathfrak{A}$ in $\mathbb{Z}[x_1,\ldots,x_n]/\mathfrak{a} $, 
and the value of maximum substitution of $\alpha$ under the isomorphism described by Equation~\eqref{isomorphism}. This
will help us to prove that $SPP$ is polynomially reducible to $SSub$ as 
the problem of finding an element with the smallest norm in an ideal, $\mathfrak{A}$ in 
$\mathbb{Z}[x_1,\ldots,x_n]/\sqrt{\mathfrak{a}}$ is equivalent to the problem of finding an element $\alpha$ in the ideal $\psi^{-1}(\mathfrak{A})$ in $\mathbb{Z}[a_1,\ldots,a_n]$
with the smallest $\mathrm{maxCoeff}_{(a_1, \ldots, a_n)}(\alpha)$. 

The following result is easy to see.
\begin{lemma}\label{lowerbound}
 Let $\mathfrak{a} \subseteq \mathbb{Z}[x_1,\ldots,x_n]$ be an ideal such that  $\mathbb{Z}[x_1,\ldots,x_n]/\mathfrak{a} $ 
 is finitely generated and has a free $\mathbb{Z}$-module representation w.r.t. a monomial order, $\prec$. Let the finite set of zeroes, $\mathcal{V}(\mathfrak{a})$ be of cardinality $N$. Let $\mathcal{B}$ be the canonical basis of the free residue class ring 
  constructed using \citep[Theorem 4.1]{FrancisDukkipati:2013:freeZmodule}.  Let  $\alpha \in \mathbb{Z}[a_1, \ldots, a_n]$, $(a_1, \ldots, a_n) \in \mathcal{V}(\mathfrak{a})$. Let  $\psi$ be the isomorphism given by Equation~\eqref{isomorphism} and 
  corresponding to each element in $\mathcal{V}$ we  have $\psi_{i}$, $1\leq i \leq N$.
 Let $t=
  \max\limits_{x^\beta \in \mathcal{B}}(\mathrm{maxsub}(\psi^{-1}(x^\beta))). $
 Then, 
 \begin{displaymath}
  \mathrm{maxsub}(\alpha) \leq Nt\hspace{2pt}\mathrm{maxCoeff}_{({a_1}^{(i)}, \ldots, {a_n}^{(i)})}(\alpha),
 \end{displaymath}
 where $({a_1}^{(i)}, \ldots, {a_n}^{(i)})$ corresponds to $\psi_{i}$, $i  = 1, \ldots, N$. 
\end{lemma}
The above result allows us to upper bound the maximum substitution w.r.t. a factor (polynomial in $N$) of the maximum coefficient.
To prove that $SPP$  can be polynomially reduced to 
$SSub$ and vice-versa, we need to give an upper bound for the maximum coefficient
w.r.t. the maximum substitution value. We first give a result that upper bounds the maximum coefficient value to a factor (that is not a polynomial in $N$) of the 
maximum substitution value.  Then for the specific case of 
\begin{displaymath}
\mathfrak{a} = \langle {x_1}^{r_1 - 1} + {x_1}^{r_1 - 2}+ \cdots +1, \ldots, {x_n}^{r_n -1}+{x_n}^{r_n -2}+ \cdots + 1\rangle,
\end{displaymath}
we give an upper bound to a factor of $N$. 
\begin{lemma}\label{onebound}
 Let $G$ be a short reduced Gr\"obner basis of an ideal $\mathfrak{a} \subseteq \mathbb{Z}[x_1,\ldots,x_n]$  such that  $\mathbb{Z}[x_1,\ldots,x_n]/\mathfrak{a} $ 
 is  finitely generated and has a free $\mathbb{Z}$-module representation w.r.t. $G$.
 Let the finite set of zeroes, $\mathcal{V}(\mathfrak{a})$ be of cardinality $N$ and $\mathcal{B}$ be the canonical basis of the free residue class ring. 
  Let  $\alpha \in \mathbb{Z}[a_1, \ldots, a_n], (a_1, \ldots, a_n) \in \mathcal{V}(\mathfrak{a})$. 
 We have $\mathrm{maxsub}(\alpha) \in \mathbb{C}$. We denote the 
 $\max\limits_{x^\beta \in \mathcal{B}}(\mathrm{maxsub}(\psi^{-1}(x^\beta)))$
 by $t$.
 Let $\psi_{i}$ be the $N$ distinct isomorphisms in Equation~\eqref{isomorphism} for each element in $\mathcal{V}$. For  
$ i \in \{1, \ldots,n\}$,  let 
 \begin{displaymath} r_i = \max\{\nu : \nu \in \mathbb{N}, \mathrm{lt} (g) = {x_i}^\nu, g\in G \}.
 \end{displaymath}
  Suppose the following conditions are satisfied.
 \begin{enumerate}[(1)]
  \item There exists an integer tuple $(m_1,\ldots, m_n)$, $m_i \in \mathbb{N}$, $m_i \geq r_i$ such that
  for all $1 \leq k \leq N$ and for $(j_1, \ldots, j_n)$ such that $j_i \leq m_i - 1$ we have,
  \begin{enumerate}[(a)]
  \item $1\leq \Big |{{a_1}^{(k)}}^{j_1} \ldots  {{a_n}^{(k)}}^{j_n} \Big| \leq  t$ and
  \item for every $({a_1}^{(k)}, \ldots, {a_n}^{(k)}) \in \mathcal{V}(\mathfrak{a})$,  
  \begin{displaymath}
   \sum_{k=1}^N {({a_1}^{(k)})}^{m_1}\ldots {({a_n}^{(k)})}^{m_n} \geq N
  \end{displaymath}
  \end{enumerate}
  \item There exists a constant $s$ such that for all $(j_1, \ldots, j_n)$, where $j_i \neq 0 \hspace{2pt} \mathrm{mod}\hspace{2pt} m_i$ and for $k \in \{1, \ldots, N\}$,
we have,
\begin{displaymath}
 \Big | \sum_{k=1}^N {({a_1}^{(k)})}^{j_1}\ldots {({a_n}^{(k)})}^{j_n} \Big|\leq s \leq 1.
\end{displaymath}
 \end{enumerate}
Then for all $\alpha \in \mathbb{Q}$, we have 
\begin{displaymath}
\mathrm{maxCoeff}_{({a_1}^{(1)}, \ldots, {a_n}^{(1)})} (\alpha) \leq \Big (\frac{Nt}{N(1-s) + s} \Big )\mathrm{maxsub}(\alpha).
\end{displaymath}
\end{lemma}
\proof
The existence of $r_i, i = 1,2,\ldots, n$ is assured by \cite[Theorem 4.3]{FrancisDukkipati:2013:freeZmodule}. For each $(j_1, \ldots, j_n)$ such that $0\leq j_i\leq r_i - 1$, 
we have the following set of $N$ inequalities, $1 \leq k \leq N$
\begin{displaymath}
\Big | \psi_k (\alpha) {{a_1}^{(k)}}^{m_1 - r_1 + j_1} \cdots  {{a_n}^{(k)}}^{m_n - r_n + j_n} \Big| \leq  \mathrm{maxsub}(\alpha)t.
\end{displaymath}
This is because by definition $ | \psi_k (\alpha)  | \leq  \mathrm{maxsub}(\alpha)$ and by  (1.a),
\begin{displaymath}
 |{{a_1}^{(k)}}^{m_1 - r_1 + j_1} \ldots  {{a_n}^{(k)}}^{m_n - r_n + j_n} | \leq  t. 
\end{displaymath}
We look at the the system of inequalities for a specific $(j_1, \ldots, j_n)$.  Let  $\alpha = \sum_{i=1}^{N}{\alpha_{(i_1,\ldots,i_n)}{a_1}^{i_1}\cdots{a_n}^{i_n}}$.
We have, 
\begin{displaymath}
\psi_{j}(\alpha) =  \sum_{i=1}^{m}{\alpha_{(i_1,\ldots,i_n)}{{a_1}^{(j)}}^{i_1}\ldots{{a_n}^{(j)}}^{i_n}},
\end{displaymath} where $({a_1}^{(j)},\ldots, {a_n}^{(j)})\in \mathcal{V}(\mathfrak{a})$. For $k \in \{1, \ldots, N\}$ we have,
\begin{align*}
 &\Big |\psi_k(\alpha)({{a_1}^{(k)}}^{m_1-r_1+j_1}\ldots {{a_n}^{(k)}}^{m_n-r_n+j_n})\Big | \\
 & = \Big|\alpha_{(0,0,\ldots,0)}{{a_1}^{(k)}}^{m_1-r_1+j_1}\cdots {{a_n}^{(k)}}^{m_n-r_n+j_n}+
 \cdots \\
 &\hspace{12pt}+\alpha_{(r_1-j_1,\ldots,r_n-j_n)}{{a_1}^{(k)}}^{m_1}\ldots {{a_n}^{(k)}}^{m_n}+ 
 \cdots+\alpha_{(r_1-1,\ldots,r_n - 1)}{{a_1}^{(k)}}^{m_1+j_1-1}\ldots {{a_n}^{(k)}}^{m_n+j_n-1}\Big|\\ 
 &\leq \mathrm{maxsub}(\alpha)t.
\end{align*}
Let $A = \sum_{i=1}^N {\alpha_{(i_1,\ldots,i_n)}}$ and $S_{(j_1,\ldots, j_n)} = \sum_{i=1}^{N} {{a_1}^{(i)}}^{m_1 - r_1 + j_1}\cdots{{a_n}^{(i)}}^{m_n - r_n + j_n}$.  
Then, 
\begin{align*}
&N | \alpha_{{r_1 - j_1}, \ldots, {r_n - j_n}} |- s(A - |
\alpha_{{r_1 - j_1}, \ldots, {r_n - j_n}} |)\\
&= N | \alpha_{{r_1 - j_1}, \ldots, {r_n - j_n}}  |- s(|\alpha_{(0,\ldots,0)}|+ 
\cdots + 
|\alpha_{(r_1-j_1-1,\ldots,r_n-j_n-1)}| \\ 
&\hspace{12pt}+ |\alpha_{(r_1-j_1+1,\ldots,r_n-j_n+1)}|+
\cdots + |\alpha_{(r_1-1,\ldots,r_n-1)}|)\\
&\leq |\alpha_{(r_1-j_1, \ldots, r_n-j_n)}S_{(r_1,\ldots,r_n)}| - (|\alpha_{(0,\ldots,0)}S_{(j_1,\ldots,j_n)}|+ 
\cdots + |\alpha_{(r_1-j_1-1,\ldots,r_n-j_n-1)}S_{(r_1-1,\ldots,r_n-1)}|+\\
&\hspace{16pt} |\alpha_{(r_1-j_1+1,\ldots,r_n-j_n+1)}S_{(r_1+1,\ldots,r_n+1)}|+ 
\cdots + |\alpha_{(r_1-1+j_1,\ldots,r_n-1+j_n)}|)\\
&\leq |\psi_{1}(\alpha) {{a_1}^{(1)}}^{m_1 - r_1 + j_1}\cdots{{a_n}^{(1)}}^{m_n - r_n + j_n}|+ 
\cdots + |\psi_{N}(\alpha) {{a_1}^{(N)}}^{m_1 - r_1 + j_1}\cdots{{a_n}^{(N)}}^{m_n - r_n + j_n}|\\
&\leq Nt \hspace{2pt} \mathrm{maxsub}(\alpha).
\end{align*}
This implies,
\begin{align*}
 (N + s) |  \alpha_{{r_1 - j_1}, \ldots, {r_n - j_n}} | - sA &\leq Nt \hspace{2pt} \mathrm{maxsub}(\alpha) \\
  |  \alpha_{{r_1 - j_1}, \ldots, {r_n - j_n}} | &\leq  \frac {Nt \hspace{2pt} \mathrm{maxsub}(\alpha) + sA} {N+s}.
\end{align*}
Let $B = \frac{Nt\hspace{2pt}\mathrm{maxsub}(\alpha) + sA}{N+s}$. Since $A =   \sum_{i=1}^N {\alpha_{(i_1,\ldots,i_n)}}$ we get 
$A \leq N\times B$.  
We have,
\begin{displaymath}
(N + s - ns)B \leq Nt\hspace{2pt}\mathrm{maxsub}(\alpha).
\end{displaymath}
We have $ |  \alpha_{{r_1 - j_1}, \ldots, {r_n - j_n}} | \leq B$, which implies,
\begin{displaymath}
\mathrm{maxCoeff}_{({a_1}^{(1)}, \ldots, {a_n}^{(1)})} (\alpha) \leq \frac{Nt}{N(1-s) + s} \mathrm{maxsub}(\alpha).
\end{displaymath}
\endproof
The above lemma gives the bound that is similar to the univariate case.
We now study the above lemma for the specific case of 
\begin{displaymath}
\mathfrak{a} = \langle {x_1}^{r_1 - 1} + {x_1}^{r_1 - 2}+ \cdots +1, \ldots, {x_n}^{r_n -1}+{x_n}^{r_n -2}+ \cdots + 1\rangle.
\end{displaymath}
In this case, $\mathrm{maxCoeff}$  is bound by a factor of $N$.
\begin{proposition}\label{variety}
 Let
 \begin{displaymath}
 \mathfrak{a} = \langle {x_1}^{r_1 - 1} + {x_1}^{r_1 - 2}+ \cdots +1, \ldots, {x_n}^{r_n -1}+{x_n}^{r_n -2}+ \cdots + 1\rangle 
 \end{displaymath} be an ideal  in $\mathbb{Z}[x_1,\ldots,x_n]$.
 Then, 
 \begin{displaymath}
 \mathcal{V}(\mathfrak{a}) = \{(a_1, \ldots, a_n) \in {\mathbb{A}_{\mathbb{C}}}^n: a_i \text{ is a zero of } {x_i}^{r_i - 1} + {x_i}^{r_i - 2}+ \cdots +1, i \in \{1, \ldots, n\}\}.
 \end{displaymath} 
\end{proposition}
\begin{proposition}\label{boundstheorem}
Let 
\begin{displaymath}
\mathfrak{a} = \langle {x_1}^{r_1 - 1} + {x_1}^{r_1 - 2}+ \cdots +1, \ldots, {x_n}^{r_n -1}+{x_n}^{r_n -2}+ \cdots + 1\rangle
\end{displaymath}
be an ideal in 
$\mathbb{Z}[x_1,\ldots,x_n]$, $\mathcal{V}$, the finite set of zeroes of cardinality $N$ and $({a_1}^{(1)},\ldots, {a_n}^{(1)})$,  one of the zeroes. 
 Let $\alpha \in \mathbb{Z}[x_1,\ldots,x_n]/\mathfrak{a} $.  
Then,
\begin{displaymath}
\mathrm{maxCoeff}_{({a_1}^{(1)},\ldots, {a_n}^{(1)})}(\alpha) \leq N \hspace{2pt}\mathrm{maxsub}(\alpha) 
\end{displaymath}
$\:\:\mathrm{and}$
\begin{displaymath}
\mathrm{maxsub}(\alpha) \leq N\hspace{2pt} \mathrm{maxCoeff}_{({a_1}^{(1)},\ldots, {a_n}^{(1)})}(\alpha).
\end{displaymath} 
\end{proposition}
\proof
By Equation~\eqref{isomorphism}, $\mathbb{Z}[x_1,\ldots,x_n]/\mathfrak{a} $ is isomorphic to $\mathbb{Z}[a_1, \ldots, a_n]$, $(a_1, \ldots, a_n) \in \mathcal{V}(\mathfrak{a})$. We have from Lemma~\ref{lowerbound} that 
\begin{displaymath}
\mathrm{maxsub}(\alpha) \leq Nt\hspace{2pt} \mathrm{maxCoeff}_{({a_1}^{(1)},\ldots, {a_n}^{(1)})}(\alpha).
\end{displaymath}
The zeroes of this ideal are the zeroes of each individual generator (Proposition~\ref{variety}). Each individual generating polynomial is a cyclotomic polynomial 
and therefore all the zeroes of generators are of norm $1$ and so we have $t=1$ and the following inequality,
\begin{displaymath}
\mathrm{maxsub}(\alpha) \leq N\hspace{2pt} \mathrm{maxCoeff}_{({a_1}^{(1)},\ldots, {a_n}^{(1)})}(\alpha).
\end{displaymath}
Now to prove that $ \mathrm{maxCoeff}_{({a_1}^{(1)},\ldots, {a_n}^{(1)})}(\alpha)\leq N \hspace{2pt}\mathrm{maxsub}(\alpha)$. 
If the conditions in  Lemma~\ref{onebound} are satisfied we have that 
\begin{displaymath}
\mathrm{maxCoeff}_{({a_1}^{(1)}, \ldots, {a_n}^{(1)})} (\alpha) \leq \frac{Nt}{N(1-s) + s} \mathrm{maxsub}(\alpha).
\end{displaymath}
 Now we show that the conditions in Lemma~\ref{onebound} are indeed satisfied. 
We have $t = 1$ and $m_i = r_i$. We need to determine if  
\begin{displaymath}
 \Big| \sum_{i=1}^{N} {{a_1}^{(i)}}^{m_1 }\ldots{{a_n}^{(i)}}^{m_n }  \Big | \geq N,
\end{displaymath}
and if we can find a $s$ such that
\begin{displaymath}
\Big | \sum_{i=1}^{N} {{a_1}^{(i)}}^{j_1 }\ldots{{a_n}^{(i)}}^{j_n }  \Big | \leq s \leq 1.
\end{displaymath}
We have that ${a_i}^{(j)}$ is the zero of ${x_i}^{r_i-1} + {x_i}^{r_i-2} + \cdots + 1$.
This implies 
\begin{displaymath}
{{a_i}^{(j)}}^{m_i} =( {{a_i}^{(j)}}{(r_i-1)} + {{a_i}^{(j)}}{r_i - 2)} + \cdots + 1) ({a_i}^{(j)} - 1) + 1 = 1.
\end{displaymath}
So,
$\Big | \sum_{i=1}^{N} {{a_1}^{(i)}}^{m_1 }\ldots{{a_n}^{(i)}}^{m_n }   \Big| = N$.
Since each generator, $g_i = {x_i}^{r_i-1} + {x_i}^{r_i-2} + \cdots + 1$, is a cyclotomic polynomial it has a zero, 
say ${{a_i}^{(1)}}$, such that all the remaining zeroes, ${{a_i}^{(j)}}$ is some power of this root, 
i.e. ${{{a_i}^{(1)}}}^{j} = {{a_i}^{(j)}} $. We also have, ${{a_i}^{(j)}}^{r_i} = 1$, $j = 1, \ldots, n$.  Therefore, 
\begin{displaymath}
{{a_i}^{(j)}}^{k} =  {{a_i}^{(j)}}^{k \hspace{2pt}\mathrm{mod}\hspace{2pt} r_i}, k \in \mathbb{N}.
\end{displaymath}
We will now  find a $s$ such that the second condition in Lemma ~\ref{onebound} is satisfied. 
For all $(j_1, \ldots, j_n)$, where $j_i \neq 0 \hspace{2pt} \mathrm{mod}\hspace{2pt} m_i$ for some $i  = 1, \ldots, n$,
we have,
\begin{equation*}
 \Big|\sum_{i=1}^m {({a_1}^{(i)})}^{j_1}\ldots {({a_n}^{(i)})}^{j_n} \Big | 
 =\Big|\sum_{i=1}^m {({a_1}^{(i)})}^{j_1 \hspace{2pt} \mathrm{mod}\hspace{2pt} m_1}\ldots {({a_n}^{(i)})}^{j_n \hspace{2pt} \mathrm{mod}\hspace{2pt} m_n} \Big |.
\end{equation*}
  We replace the zeroes with powers of ${{a_i}^{(1)}}$ for $i = 1, \ldots, n$.
  Therefore we have, 
  \begin{align*}
   &\Big | \sum_{i=1}^m {({a_1}^{(i)})}^{j_1}\ldots {({a_n}^{(i)})}^{j_n} \Big | 
   = \Big | \sum_{i=1}^m {({a_1}^{(1)})}^{{i}\hspace{2pt}{j_1 \hspace{2pt} \mathrm{mod}\hspace{2pt} m_1}}\ldots {({a_n}^{(1)})}^{{i}\hspace{2pt}{j_n \hspace{2pt} \mathrm{mod}\hspace{2pt} m_n}} \Big |\\
   & = \Big | \sum_{i=1}^m {({a_1^{({j_1 \hspace{2pt} \mathrm{mod}\hspace{2pt} m_1})})}^{i}\ldots {({a_n^{({j_n \hspace{2pt} \mathrm{mod}\hspace{2pt} m_n})})}}^{i}} \Big |
    = |-1| = 1.
  \end{align*}
 We can take $s=1$ and apply in the inequality from Lemma~\ref{onebound} to get,
\begin{displaymath}
\mathrm{maxCoeff}_{({a_1}^{(1)}, \ldots, {a_n}^{(1)})} (\alpha) \leq N \hspace{2pt} \mathrm{maxsub}(\alpha).
\end{displaymath}
\endproof
The result below connects  $SPP$ with $SSub$ by a factor that is polynomial in the cardinality of $\mathcal{V}(\mathfrak{a})$. 
\begin{theorem}
 Let 
\begin{displaymath}
\mathfrak{a} = \langle {x_1}^{r_1 - 1} + {x_1}^{r_1 - 2}+ \cdots +1, \ldots, {x_n}^{r_n -1}+{x_n}^{r_n -2}+ \cdots + 1\rangle
\end{displaymath} 
  be an ideal in 
$\mathbb{Z}[x_1,\ldots,x_n]$ . The residue class polynomial ring, $\mathbb{Z}[x_1,\ldots,x_n]/\mathfrak{a}$ is free and finitely generated.
Let  $\mathcal{V}(\mathfrak{a})$ be of cardinality $N$. Let $\psi$ represent the isomorphism
as described in Equation~\eqref{isomorphism}. Then, 
\begin{align}
&\label{Equation3} \mathfrak{a}-SPP_{\gamma{N^2}}(\mathfrak{A}) \leq
 \mathfrak{a}-SSub_{\gamma}(\psi^{-1}(\mathfrak{A}))\:\:\: \mathrm{and} \\
 &\label{Equation4}\mathfrak{a}-SSub_{\gamma{N^2}}(\psi^{-1}(\mathfrak{A})) \leq \mathfrak{a}-SPP_{\gamma}(\mathfrak{A}). 
\end{align}
\end{theorem}
\proof
Let $\psi^{-1}(\mathfrak{A}) \subseteq \mathbb{Z}[a_1, \ldots, a_n]$, $(a_1, \ldots, a_n) \in \mathcal{V}(\mathfrak{a})$,  be an ideal given by its generators $\mathcal{F}=\{f_1, \ldots, f_k\}$.
Then each element
in  $\mathcal{F}$ can be written in terms of the elements  $\{a_1,\ldots,a_n\}$ such that $(a_1,\ldots, a_n)\in \mathcal{V}$.
The oracle for  $\mathfrak{a}-SPP_{\gamma}(\mathfrak{A}))$ finds us an
element $h \in \mathfrak{A}$ such that its norm is less than $\gamma
{\lambda_1}^\infty(\mathfrak{A})$. Let $\alpha = \psi^{-1}(h)$. 
We have,
\begin{displaymath}
 \mathrm{maxCoeff}_{({a_1}, \ldots, {a_n})} (\alpha) \leq \gamma\cdot \mathrm{maxCoeff}_{({a_1}, \ldots, {a_n})} (\alpha^{'}),
\end{displaymath}
for all $\alpha^{'} \in \psi^{-1}(\mathfrak{A})$.
Applying Proposition~\ref{boundstheorem} twice we get,
\begin{align*}
 \mathrm{maxsub}&(\alpha) \leq N\cdot \mathrm{maxCoeff}_{(a_1,\ldots, a_n)}(\alpha),\\
 &\leq N\gamma \cdot \mathrm{maxCoeff}_{({a_1}, \ldots, {a_n})} (\alpha^{'}), \text{for all }\alpha^{'} \in \psi^{-1}(\mathfrak{A}),\\
 &\leq N^2 \gamma\cdot  \mathrm{maxsub}(\alpha^{'}), \; \text{for all }\alpha^{'} \in \psi^{-1}(\mathfrak{A}).
\end{align*}
Thus we have a $\gamma\cdot N^2$ approximation for $\mathfrak{a}-SSub$. Hence Equation~\eqref{Equation4} holds.

Next, we show Equation~\eqref{Equation3} holds.
The oracle for $\mathfrak{a}-SSub_{\gamma}(\psi^{-1}(\mathfrak{A}))$ finds an element $\alpha \in \psi^{-1}(\mathfrak{A})$ such that 
$\mathrm{maxsub}(\alpha) \leq \gamma \cdot\mathrm{maxsub}(\alpha^{'}), \hspace{5pt} \text{for all }\alpha^{'} \in \psi^{-1}(\mathfrak{A})$. Again we apply  Proposition~\ref{boundstheorem} twice.
\begin{align*}
 \mathrm{maxCoeff}_{(a_1,\ldots, a_n)}&(\alpha) \leq N \cdot \mathrm{maxsub}(\alpha), \\
 &\leq N\gamma \cdot \mathrm{maxsub}(\alpha^{'}), \hspace{5pt} \text{for all }\alpha^{'} \in \psi^{-1}(\mathfrak{A}),\\
 &\leq N^2 \gamma\cdot \mathrm{maxCoeff}_{(a_1,\ldots, a_n)}(\alpha^{'}),
\end{align*}
for all $\alpha^{'} \in \psi^{-1}(\mathfrak{A})$.
We have a $\gamma\cdot N^2$ approximation for $\mathfrak{a}-SPP$.
\endproof
\section{Collision Resistant Generalized Hash Functions}
\label{hashfunctionsmultivariate}
\noindent
We can construct hash function families described in Section~\ref{Preliminaries} 
based on multivariate ideal lattices. Consider a prime ideal, $\mathfrak{a} \subseteq \mathbb{Z}[x_1,\ldots,x_n]$ 
such that the residue class polynomial ring,
$\mathbb{Z}[x_1,\ldots,x_n]/\mathfrak{a}$ is free and finitely generated and is of size $N \in \mathbb{N}$.
The hash function family $\mathcal{H}(R,D,m)$ is given by
$R = \mathbb{Z}_p[x_1,\ldots,x_n]/\mathfrak{a}$, where $p\in \mathbb{N}$ is approximately of the order $N^2$ and $D$ is a strategically chosen subset of $R$ 
 and $m \in \mathbb{N}$. Let the expansion factor, $\mathcal{E}(\mathfrak{a},\prec, (3,3,\ldots,3)) \leq \eta$, for some $ \eta \in \mathbb{R}$. 
 Let $D = \{g \in R : \lVert g \rVert_{\mathfrak{a},\prec} \leq d \}$ for some positive integer $d$. Then $\mathcal{H}$ maps elements from $D^m$ to $R$. 
 We have $|D^m| = (2d+1)^{Nm}$ and  $|R| = p^N$. If $m \gneq \frac{\log p}{\log 2d} $, then $\mathcal{H}$ will
 have collisions. We show that finding a collision for a hash function randomly chosen from $\mathcal{H}$ is as hard as solving
 $\mathfrak{a}-SPP_\gamma$ for a particular ideal in $\mathfrak{A} \subseteq \mathbb{Z}[x_1,\ldots,x_n]/\mathfrak{a}$. 
As we mentioned before, even though the hardness results of univariate and multivariate ideal lattices
are based on different problems,  other properties like  collision resistance of hash functions are exactly 
analogous. 
The reader can 
 refer to \citep{Lyubashevsky:2006:Ideallatticefirstdef} for detailed constructions.
 %
 \begin{theorem}
  Consider an ideal $\mathfrak{a} \subseteq \mathbb{Z}[x_1,\ldots,x_n]$ 
such that the residue class polynomial ring,
$\mathbb{Z}[x_1,\ldots,x_n]/\mathfrak{a}$ is  finitely generated of size $N \in \mathbb{N}$ and has a free $\mathbb{Z}$-module representation w.r.t. $\prec$.
Let $\mathcal{H}(R,D,m)$ be the associated  
hash function family as mentioned above with
$R = \mathbb{Z}_p[x_1,\ldots,x_n]/\mathfrak{a}$, $m \gneq \frac{\log p}{\log 2d} $ and
  $p \geq 8\eta dmN^{1.5}\sqrt{\log N}$. Then, for $\gamma = 8\eta^2dmN\log^2N$,
  there is a polynomial time reduction from $\mathfrak{a}-SPP_\gamma(\mathfrak{A})$, for any ideal $\mathfrak{A} \subseteq \mathbb{Z}[x_1,\ldots,x_n]/\mathfrak{a}$,
  to $Collision_{\mathcal{H}}(\mathfrak{h})$ where $\mathfrak{h}$ is chosen uniformly at random from $\mathcal{H}$.  
 \end{theorem}
$Collision_{\mathcal{H}}(\mathfrak{h})$ is the problem of finding a collision given a hash function, $\mathfrak{h}$. 
The idea  is that if one can solve in polynomial time the problem $Collision_{\mathcal{H}}(\mathfrak{h})$ for a randomly chosen $\mathfrak{h}$ 
then we can solve the $\mathfrak{a}-IncSPP_\gamma$ problem for any ideal $\mathfrak{A}$ and $\gamma =8\eta^2dmN\log^2N$. 
This implies we have a polynomial reduction from $\mathfrak{a}-SPP_\gamma$ to  $Collision_{\mathcal{H}}(\mathfrak{h})$. 

We consider an oracle $\mathcal{C}$, which when given an $\mathfrak{h}$ returns a collision with non-negligible probability and in polynomial time. 
We are given an ideal $\mathfrak{A} \subseteq \mathbb{Z}[x_1,\ldots,x_n]/\mathfrak{a}$ and 
an element of the ideal $g$ such that $\lVert g \rVert_{\infty} \gneq 8\eta^2 dmN\log^2N{\lambda_1}^\infty (\mathfrak{A})$. 
We have to find a non-zero $h \in \mathfrak{A}$ such that $\lVert h \rVert_{\mathfrak{a},\prec }\leq \lVert g \rVert _{\mathfrak{a},\prec}/2$. 

Given vectors $c,x \in \mathbb{R}^N$ and any $l\gneq 0$, $\rho_{l,c}(x) = e^{-\pi{\lVert (x-c)/l\rVert}^2}$ represents a Gaussian function that has its center at $c$ and is scaled by $l$.   
The total measure is $\int_{x\in \mathbb{R}^N} \rho_{l,c}(x)dx = l^N$ and therefore $\rho_{l,c}/l^N$
is a probability density function.  \cite{Micciancio:2004:Gaussian} introduced certain techniques to approximate the distribution efficiently, 
effectively allowing us  to sample from the distribution, $\rho_{l,c}/l^N$ exactly. 
In this paper, the results are used in the same way as in \citep{Lyubashevsky:2006:Ideallatticefirstdef} as the results are for integer lattices in general and not specifically for ideal lattices in one 
variable.  

Let
$ s = \frac{\lVert g \rVert_{\infty}}{8\eta\sqrt{N} dm\log N}$. Therefore, $\lVert g \rVert_\infty = 8\eta dms\sqrt{N}\log N$. 
Also the results from \cite[Lemma 4.1]{Micciancio:2004:Gaussian} imply that if we sample $y\in\mathbb{R}^N$ from the distribution ${\rho_s}/s^N$, then 
\begin{displaymath}
\bigtriangleup(y+\mathfrak{A}, U(\mathbb{R}^N/\mathfrak{A}))\leq (\log N)^{-2\hspace{2pt}\log N}/2,
\end{displaymath}
 i.e.  $y + \mathfrak{A}$ is a uniformly random coset.
We list a procedure in Algorithm~1, by which using the
access to the oracle one can determine an $h$ such that it is a
solution to the $IncSPP_\gamma$ problem.  
\begin{algorithm}\label{Algorithm}
\caption{Finding the solution of the $IncSPP_\gamma$ problem given access to the $Collision$ oracle } 
\begin{algorithmic}[1]
\STATE \textbf{Input} Finitely generated $\mathbb{Z}[x_1,\ldots,x_n]/\mathfrak{a}$ with a free $\mathbb{Z}$-module representation w.r.t. $\prec$, \\
$\mathfrak{A} \subseteq \mathbb{Z}[x_1,\ldots,x_n]/\mathfrak{a}$ an ideal, and\\
$g \in \mathfrak{A}$ such that $\lVert g \rVert_\infty = 8\eta dms\sqrt{N}\log N$.
\STATE \textbf{Output} $h \in \mathfrak{A}$ such that $\lVert h \rVert \neq 0$ $\lVert h \rVert_{\mathfrak{a},\prec }\leq \lVert g \rVert _{\mathfrak{a},\prec}/2$.
\FOR {$i = 1$ to $m$}
\STATE Generate a random coset of $\mathfrak{A}/\langle g \rangle$ and let $v_i$ be a polynomial in that coset.
\STATE Generate $y_i \in \mathbb{R}^N$ such that $y_i$ has distribution $\rho_s/s^n$ and consider $y_i$ as a polynomial in $\mathbb{R}[x_1,\ldots,x_n]$. 
\STATE Let $w_i \in \mathbb{R}[x_1,\ldots,x_n]$ be the unique polynomial such that $p(v_i+y_i) \equiv gw_i$ in $\mathbb{R}^N/\langle pg\rangle$.
Note that the coefficients of $w_i$ lie in $[0,p)$.
\STATE Let $a_i = [w_i] \mathrm{mod} \hspace{2pt} p$ .
\ENDFOR
\STATE Give $(a_1,\ldots,a_m)$ as input to the oracle $C$ and using its output determine polynomials $z_1, \ldots, z_m$ such that $\lVert z \rVert_{\mathfrak{a},\prec} \leq 2d$ and
$\sum_{i=1}^{m} z_ia_i \equiv 0$ in the ring $\mathbb{Z}_p[x_1,\ldots,x_n]/\mathfrak{a}$. (Details of the construction of $z_i$ can be found in Lemma~\ref{lemma5.2}).
\STATE Output $h = \Big(\sum_{i=1}^{m}\Big(\frac{g(w_i - [w_i]}{p} - y_i\Big)z_i\Big) \mathrm{mod}\hspace{2pt} \mathfrak{a}$.
\end{algorithmic}
\end{algorithm}
Now, it is enough to show that  Algorithm~1 runs in
polynomial time, the inputs to the oracle are uniformly random, and
$h$ satisfies all the desired properties. 
\begin{lemma}\label{lemma5.2}
 Algorithm 1 runs in polynomial time.
\end{lemma}
\proof
In Step (4), we need to generate a random coset of  $\mathfrak{A}/\langle g \rangle$. Since $\mathfrak{a}$ is a prime ideal, the ideals $\mathfrak{A}$ and $\langle g \rangle $ are $\mathbb{Z}$-modules of dimension $n$.  
There is a polynomial time algorithm to generate a random element from $\mathfrak{A}/\langle g \rangle$ \cite[Proposition 8.2]{Micciancio:2002:CyclicLattices}. Step (5) and Step (6) will be justified in the following lemma.
Step (7) just rounds off the coefficients and  takes modulo $p$ and therefore can be done in polynomial time. 
In Step (9), we feed $(a_1, \ldots, a_m)$ to the $Collision$ oracle and it returns $(\alpha_1, \ldots, \alpha_m), (\beta_1,\ldots, \beta_m)$ 
such that $\lVert \alpha_i \rVert_{\mathfrak{a},\prec}, \lVert \beta_i \rVert_{\mathfrak{a},\prec} \leq d$ and 
$\sum_{i=1}^m a_i\alpha_i \equiv  \sum_{i=1}^m a_i\beta_i $ in $\mathbb{Z}_p[x_1,\ldots,x_n]/\mathfrak{a}$. 
Therefore, if we set $z_i = \alpha_i - \beta_i$,  it satisfies the properties of Step (9).
\endproof
\begin{lemma}
Consider the polynomials $a_i$ as elements in ${\mathbb{Z}_p}^N$. Then,
\begin{displaymath}
\bigtriangleup((a_1,\ldots,a_m), U({\mathbb{Z}_p}^{N\times m})) \leq m(\log N)^{-2\log N}/2.
\end{displaymath}
\end{lemma}
\proof
We have chosen $v_i$ from a uniformly random coset of $\mathfrak{A}/\langle g \rangle$. If $y_i$ is in a uniformly random coset of $\mathbb{R}^N/\langle g \rangle$, then $p(v_i + y_i)$ is a uniformly random coset of $\mathbb{R}^N/\langle pg \rangle$. A basis for $\mathbb{R}^N/\langle pg \rangle$ is $\{pgb_1, \ldots, pgb_N\}$ where $\{ b_1, \ldots, b_N\}$ is the basis of $\mathbb{Z}[x_1,\ldots, x_n]/\mathfrak{a}$. 
Every element in $\mathbb{R}^N/\langle pg \rangle$ can be represented as $\alpha_0pgb_1 +  \cdots + \alpha_Npgb_N$ where $\alpha_i \in [0,1)$.
 Therefore Step (6) is justified with $w_i = \alpha_0pb_1 +  \cdots + \alpha_Npb_N$. Since we have assumed $p(v_i + y_i)$ is a uniformly random coset of $\mathbb{R}^N/\langle pg \rangle$ the coefficients of $w_i$ are uniform over $[0,p)$ and the input to the oracle in Step (9) is correct. 
 The only thing remaining is to check if the assumption that $y_i$ is in a uniformly random coset of $\mathbb{R}^N/\langle g \rangle$ is correct. It is not exactly uniformly random but very close to it. We have 
 $\bigtriangleup(\rho_s/s^n+\mathfrak{A}, U(\mathbb{R}^N/\mathfrak{A}))\leq (\log N)^{-2\log N}/2$.  Since $a_i$ is a function of $y_i$, we have $\bigtriangleup(a_i, U({\mathbb{Z}_p}^N)) \leq (\log N)^{-2\log N}/2$. Since all the $a_i$s are independent we have 
   $\bigtriangleup((a_1,\ldots,a_m),U({\mathbb{Z}_p}^{N\times m})) \leq m(\log  N)^{-2\log N}/2$.
\endproof

The following three lemmas ensure that the output of the algorithm, $h$  satisfies the desired properties of the $IncSPP_\gamma$ problem, i.e. 
$h$ is non zero, $h\in \mathfrak{A}$ and  $\lVert h \rVert_{\mathfrak{a},\prec} \leq \frac{\lVert g \rVert_\infty}{2} $. 
\begin{lemma}
$h\in \mathfrak{A}$.
\end{lemma}
\proof
The proof proceeds exactly in the same lines as the univariate case. See \cite[Lemma 5.4]{Lyubashevsky:2006:Ideallatticefirstdef}.
\endproof

\begin{lemma}
With probability negligibly different from $1$, $\lVert h \rVert_{\mathfrak{a},\prec} \leq \frac{\lVert g \rVert_\infty}{2}$.
\end{lemma}
\proof
See  proof of \cite[Lemma 5.5]{Lyubashevsky:2006:Ideallatticefirstdef}.
\endproof

\begin{lemma}
$Pr[h\neq 0|(a_1,\ldots,a_m)(z_1,\ldots,z_m)]=\Omega(1)$.
\end{lemma}
\proof
See proof of \cite[Lemma 5.6]{Lyubashevsky:2006:Ideallatticefirstdef}.
\endproof
\section{Concluding remarks}
\label{conclusion}
\noindent
In this paper, we study ideal lattices in the multivariate case
and show how short reduced Gr\"obner bases can be used to locate them. 
We show that ideal lattices in the multivariate case are a generalization of
multivariate cyclic lattices, thus drawing parallels with univariate ideal lattices. 
We also
 provide a necessary and sufficient condition for full rank ideal lattices.
We establish the existence of  generalized hash functions based
on multivariate ideal  lattices and prove that they are indeed collision resistant. This
class of generalized hash functions includes hash functions  based on univariate 
ideal lattices that were previously studied in cryptography. We propose
certain worst case problems based on which we establish the security
of these hash functions. 
We show the hardness of these problems for
$\mathfrak{a} = \langle {x_1}^{r_1 - 1} + {x_1}^{r_1 - 2}+ \cdots +1,
\ldots, {x_n}^{r_n -1}+{x_n}^{r_n -2}+ \cdots + 1\rangle$. A possible future direction
is to determine the hardness of these problems for other choices of
$\mathfrak{a}$.  

Unlike in the univariate case, here  we cannot bound the expansion
factor tightly because both the structure of the ideal and  the polynomial being reduced  have a role to play in the
number of iterations in the reduction. In the univariate case
 an intuition can be given on
how to select an ideal with a ``small" expansion factor \citep{Lyubashevsky:2006:Ideallatticefirstdef}. It would be
an interesting problem to come up with similar observations in the
multivariate case.  
Polynomial computations in the univariate case are well studied and efficient methods using FFT have been proposed.
A major challenge for practical implementations using multivariate ideal lattices is coming up with similar efficient methods for multivariate  polynomial 
computations. 
We also need to study the security issues of  multivariate ideal lattices.
Another interesting direction is to see if other cryptographic
primitives like digital signatures, identification schemes can be
built from multivariate ideal lattices.  

\section*{Acknowledgments}
\footnotesize{
The authors would like to thank Debarghya Ghoshdastidar for useful
discussions on tensor representations of cyclic lattices in the
multivariate case.}

\end{document}